\documentclass[twocolumn]{aastex62}

\accepted{November 25, 2018}
\submitjournal{AJ}

\begin{document}
\title{Estimating Planetary Mass with Deep Learning}

\correspondingauthor{Elizabeth J. Tasker}
\email{elizabeth.tasker@jaxa.jp}

\author[0000-0001-6692-612X]{Elizabeth J. Tasker}
\affil{Institute of Space and Astronautical Science, Japan Aerospace Exploration Agency, \\
Yoshinodai 3-1-1, Sagamihara, Kanagawa 252-5210, Japan}

\author[0000-0001-6022-0046]{Matthieu Laneuville}
\affiliation{Earth-Life Science Institute, Tokyo Institute of Technology, 2-12-1 Tokyo, Japan}

\author{Nicholas Guttenberg}
\affiliation{Earth-Life Science Institute, Tokyo Institute of Technology, 2-12-1 Tokyo, Japan}
\affiliation{Araya Inc., Toranomon 15 Mori Building 2F, 2-8-10 Toranomon, Minato-ku, Tokyo 105-0001, Japan}

\begin{abstract}
While thousands of exoplanets have been confirmed, the known properties about individual discoveries remain sparse and depend on detection technique. To utilize more than a small section of the exoplanet dataset, tools need to be developed to estimate missing values based on the known measurements. Here, we demonstrate the use of a neural network that models the density of planets in a space of six properties that is then used to impute a probability distribution for missing values. Our results focus on planetary mass which neither the radial velocity nor transit techniques for planet identification can provide alone. The neural network can impute mass across the four orders of magnitude in the exoplanet archive, and return a distribution of masses for each planet that can inform about trends in the underlying dataset. The average error on this mass estimate from a radial velocity detection is a factor of 1.5 of the observed value, and 2.7 for a transit observation. The mass of Proxima Centauri b found by this method is $1.6^{\rm +0.46}_{\rm -0.36}$\,M$_\oplus$, where the upper and lower bounds are derived from the root mean square deviation from the log mass probability distribution. The network can similarly impute the other potentially missing properties, and we use this to predict planet radius for radial velocity measurements, with an average error of a factor 1.4 of the observed value. The ability of neural networks to search for patterns in multidimensional data means that such techniques have the potential to greatly expand the use of the exoplanet catalogue.
\end{abstract}


\section{Introduction} 
\label{sec:intro}

Since the discoveries of the first extrasolar planets in the early 1990s, the number of confirmed planets beyond our Solar System has exploded in a rise of discoveries that shows no signs of abating. The exoplanet catalog now contains over 4000 confirmed entries\footnote[1]{ \url{https://exoplanetarchive.ipac.caltech.edu/}}, a substantial fraction of which were found using dedicated planet hunters, such as the Kepler Space Telescope and the Convection, Rotation et Transits plan\'{e}taires (CoRoT). These numbers are anticipated to continue to boom through current and near-future instruments such as the Transiting Exoplanet Survey Satellite (TESS), CHaracterising ExOPlanet Satellite (CHEOPS) and PLAnetary Transits and Oscillations of stars (PLATO), Japan Astrometry Satellite Mission for INfrared Exploration (JASMINE), Gaia, as well as ground-based counterparts. 

Yet, the measured properties of individual exoplanets remain limited and dependent on the detection technique used in each particular discovery. 96\% of exoplanets have been identified using either the radial velocity or the transit technique. The former measures the radial motion of the host star due to the planet's gravity to give the planet's orbital period, orbital eccentricity and minimum mass. The true mass requires knowledge of the orbital inclination of the planet, which cannot be measured through the stellar radial velocity alone. The transit technique measures the dip in brightness of a planet passing in front of the host star as seen from Earth. This provides information on the orbital period, inclination angle and planet radius. Stellar data is also available but may be incomplete, including properties such as stellar mass, radius and luminosity. The result is a large but sparse dataset for exoplanets.

This situation is unlikely to change substantially with future observations. Many (most likely most) planets do not transit, with the geometric probability decreasing as the inverse of the orbital radius, $p_T = R_*/a$ for stellar radius $R_*$ and orbital radius, $a$. An Earth-sized planet orbiting a Sun-like star on our current orbit would therefore have just 0.47\% chance of transiting and allowing the properties associated with that detection to be recorded. The most common orbital period of currently known transiting planets is between 2 - 6 days\footnote[1]{\url{https://exoplanetarchive.ipac.caltech.edu/}}, boding ill for exploring planetary system architectures similar to our own through this method. Likewise, stellar activity often excludes a radial velocity detection that could provide an eccentricity or minimum mass \citep{Oshagh2017}. 

The problem this gives can be easily demonstrated by our nearest neighbor. Proxima Centauri is a red dwarf star at a distance of 1.295 parsecs that is proposed to be a distant member of a triple system with stars $\alpha$ Centauri A \& B. High precision radial velocity measurements of the star revealed the presence of an orbiting planet with a period of 11.186 days (semi-major axis of 0.0485\,AU) and minimum mass of M$_{\rm p}\sin i = 1.27$\,M$_\oplus$, where M$_{\rm p}$ is the true mass of the planet and $i$ is the unknown orbital inclination angle \citep{ProxCenb}. With the possible exception of the passage of a free-floating rogue world, this makes Proxima Centauri b our nearest possible extrasolar planet. As Proxima Centauri has a stellar mass of just 0.12\,M$_\odot$, the planet's short orbit is within the so-called {\it habitable zone} where liquid water could exist on the planet's surface if the surface pressure and atmospheric composition were similar to that on Earth \citep{habzone}. This has resulted in Proxima Centauri b being a major focus for habitability studies and the target for prospective future observations to search for biosignatures \citep{Turbet2016, Kane2017, Luger2017, Snellen2017}. 

However, the surface conditions for Proxima Centauri b are unknown. The planet has not been observed to transit the star, leaving its orbital inclination angle unconstrained \citep{jenkins2019, Kipping2017}. In practice, this means the planet's true mass may be anything from 1.27\,M$_\oplus$ for an edge-on inclination of $i = 90^\circ$ to greater than $15$\,M$_\oplus$ for $i < 5^\circ$; a range that covers compositions consistent with rocky planets like the Earth through to Neptune-sized gas giants with a core buried deep beneath a colossal atmosphere. 

The reverse problem occurs for planets identified with the transit technique. Empirical results based on a small subset of planets with both mass and radii measurements, suggests that planets do not have thick envelopes if their radii $< 1.6$\,R$_\oplus$ \citep{rogers2015}. However, the spread in known masses for planets of a given radius extends over a factor ten \citep{Lopez2016}. This therefore provides little information to narrow down their composition. 

The next generation of telescopes that include the James Webb Space Telescope (JWST), the Atmospheric Remote-sensing Infrared Exoplanet Large-survey (ARIEL) and the European Extremely Large Telescope (E-ELT) will focus on learning far more about these worlds through atmospheric composition studies. Yet these gaps for the bulk properties of exoplanets may result in target selection for these and future missions prioritising the wrong objects. Moreover, trends to develop theories for planet formation are often drawn from small sections of the data where the necessary properties have been measured \citep{wm2014}. This is naturally unsatisfactory, as it requires discarding the majority of a large database that is extremely challenging and expensive to acquire. The alternative to this is to develop tools to estimate missing values in the catalogue. 

Many trends between planet properties have already been observed and used to impute unknown values. Hot Jupiters cluster around 1\,AU while shorter orbits are occupied by smaller planets. Cooler stars have a lower occurrence rate of larger planets, while stellar metallicity is positively correlated with the presence of gas giants and inversely so with planet number \citep{Fischer2005, Dressing2013, Fulton2018, Brewer2018}. Several attempts have also been made to correlate mass and radius for planets of particular sizes and orbits \citep{wm2014, mrforecaster}. While some of these trends have been predictable based on our understanding of planet formation, others have forced our theories to undergo radical adjustment. Connections between multiple parameters are also much harder to discover once the relationship cannot be represented on a two- or three- dimensional plot. For these reasons, we explored the use of neural networks, which allow us to model population statistics without having to impose an a priori functional form.

\begin{figure*}
    \centering
    \includegraphics[width=\textwidth]{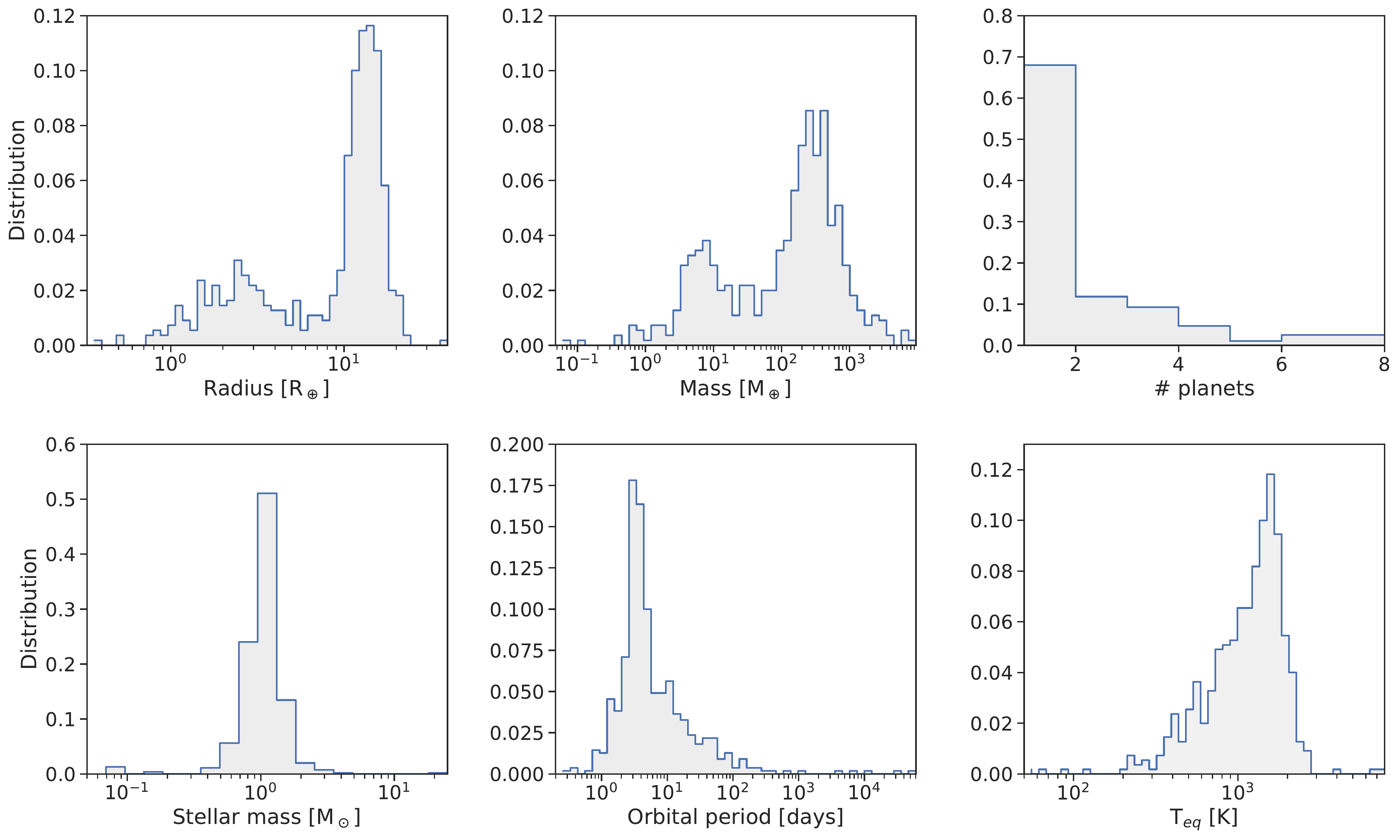}
    \caption{Distributions of the six properties in the training and test data. Top row: planet radius, planet mass, number of planets in the system. Bottom row: stellar mass, orbital period and planet equilibrium temperature.}
    \label{fig:trainingdata}
\end{figure*}

Neural networks have recently been embraced by the exoplanet community as a way of recognising features in complex or noisy data such as atmospheric spectra or a time series of light curves \citep{marquez2018, pearson2018, shallue2018}. However, they have rarely been used to contribute missing information. Here we develop a generative model that can impute a distribution for unknown exoplanet properties based on current measurements. The network models the joint probability density function of a set of planetary properties from a set of confirmed planets for which all values are known. Missing values are drawn from this density function to give an estimate consistent with the observed population. The generated values can have a complex, multidimensional dependence on the other variables, without any prior expectation on what form that dependence should take. 

This methodology differs from a Bayesian approach, where a model of the data-generating process utilises explicit prior beliefs about the model parameters. In the Bayesian approach, interpretation of the parameters is therefore with regards to the data-generating model which is tied to a set of potential claims about how the data is created. Conversely, the neural network serves a role as a universal function approximator. Specific values or distributions over coefficients of the model are not intended to have a direct physical interpretation, nor is the architecture chosen in order to express explicit prior knowledge about the process.

In this paper, we primarily focus on imputing the value for planetary mass. The mass of a planet is key to any argument related to composition or formation history \citep{unterborn2018, bond2010}, making it perhaps the most essential single measurement for understanding planetary diversity. Measuring the mass is usually achieved via a dual detection of the planet through both radial velocity and transit methods or through variations in the timing of the transit due to the gravitational pull of neighbouring planets. While direct imaging can also provide a mass, this technique is best suited to very distant, young gas giant worlds and the estimates are strongly dependent on unconstrained models for the planet evolution. This makes mass a particularly challenging measurement, as either detections through multiple methods or the presence of closely packed neighbouring planets are required. It is particularly difficult for planets with orbits larger than 6 days which are less likely to transit. 

We test the accuracy of the neural network model by predicting the mass of planets with an observed value, comparing the performance when the model is passed planet properties consistent with a radial velocity and a transit detection. We extend this to the prediction of the planet radius in the case of a radial velocity detection. We then use the model to predict the mass and radii of planets without observed measurements, with a focus on those whose currently known properties suggest they may be rocky worlds orbiting within the habitable zone. We also demonstrate how to impose additional constraints on the solution to allow for extra information in addition to the planet properties or bias considerations. We compare these results both with one dimensional estimates based on a single planet property, and also with alternative density modelling techniques that can impute properties from multiple dimensions.

While the results presented here focus on estimating planetary mass and radius, the technique can be used to estimate other planetary properties with sufficient measurements for network training. This technique can therefore be expanded in the future as the exoplanet catalog continues to grow. 

In section~\ref{sec:methods} we describe the principals behind the method used by the neural network and the network architecture. Section~\ref{sec:knownmass} examines the network performance on planets with known mass and radius values first assuming detection via the radial velocity method and then with the transit technique. In section~\ref{sec:smallplanets}, mass and radius estimates from the neural network are presented for planets without those measured properties. Section~\ref{sec:densitymodels} compares the use of the neural network with two other density model methods. Section~\ref{sec:bias} looks at the issue of transmitted bias to the network, and our results are summarised in section~\ref{sec:conclusions}. The appendix in section~\ref{sec:networktest} considers the sensitivity of the network to parameter choices.

\section{Method: Imputing planet properties}
\label{sec:methods}

The neural network employed is a modified Boltzmann machine generative model \citep{Ackley1985, chen2003continuous, Hinton2012}. The network learns the joint distribution of available exoplanet properties in order to generate new data points that lie within the same distribution. The properties chosen comprise of six observable: planet mass, planet radius, orbital period, stellar mass, equilibrium temperature and the number of known planets in the system. This selection is a compromise between the quantity of information per planet with which to impute values (more being better), and the size of the resultant training set where all entries are required to have a complete set of variables (which decreases with number of properties). Our data was taken from the NASA exoplanet archive\footnote[1]{\url{https://exoplanetarchive.ipac.caltech.edu/}} and comprises of 550 entries. To ensure the largest possible data set, missing values of our six planet properties in the confirmed planet record on the exoplanet archive were compared with the associated KOI (Kepler Object of Interest) entry where available, and that value used if present. Additionally, the equilibrium temperature was calculated from the stellar radius ($R_*$), stellar effective temperature ($T_*$) and average orbital distance ($\left<a_p\right>$) via ${\rm T_{eq} = T_{*}\sqrt{R_*/(2.0\,\left<a_p\right>)}}$ where no value was entered in the catalog. The distribution of each of the six properties in the training and test datasets are shown in Figure~\ref{fig:trainingdata}.  The complete set is also available as online material in a machine-readable format. Due to ease of detection, the most common planet types in the observational data are large worlds on short orbits. The impact such observational bias has on the results and ways in which this could be tackled without discarding the majority of the observations will be discussed in section~\ref{sec:bias}.

The network uses a training set of the six properties for 400 observed exoplanets, $x$, to create a log probability function, $H(x)$, which is then sampled via Hamiltonian Monte-Carlo to generate new data, $\tilde{x}$. All six properties can be generated to create an artificial planet whose features are consistent with the observed dataset. More usefully, a single property can be generated for an observed planet whose other measured properties are fixed and left unchanged by the network.

\begin{figure}
    \centering
    \includegraphics[width=\columnwidth]{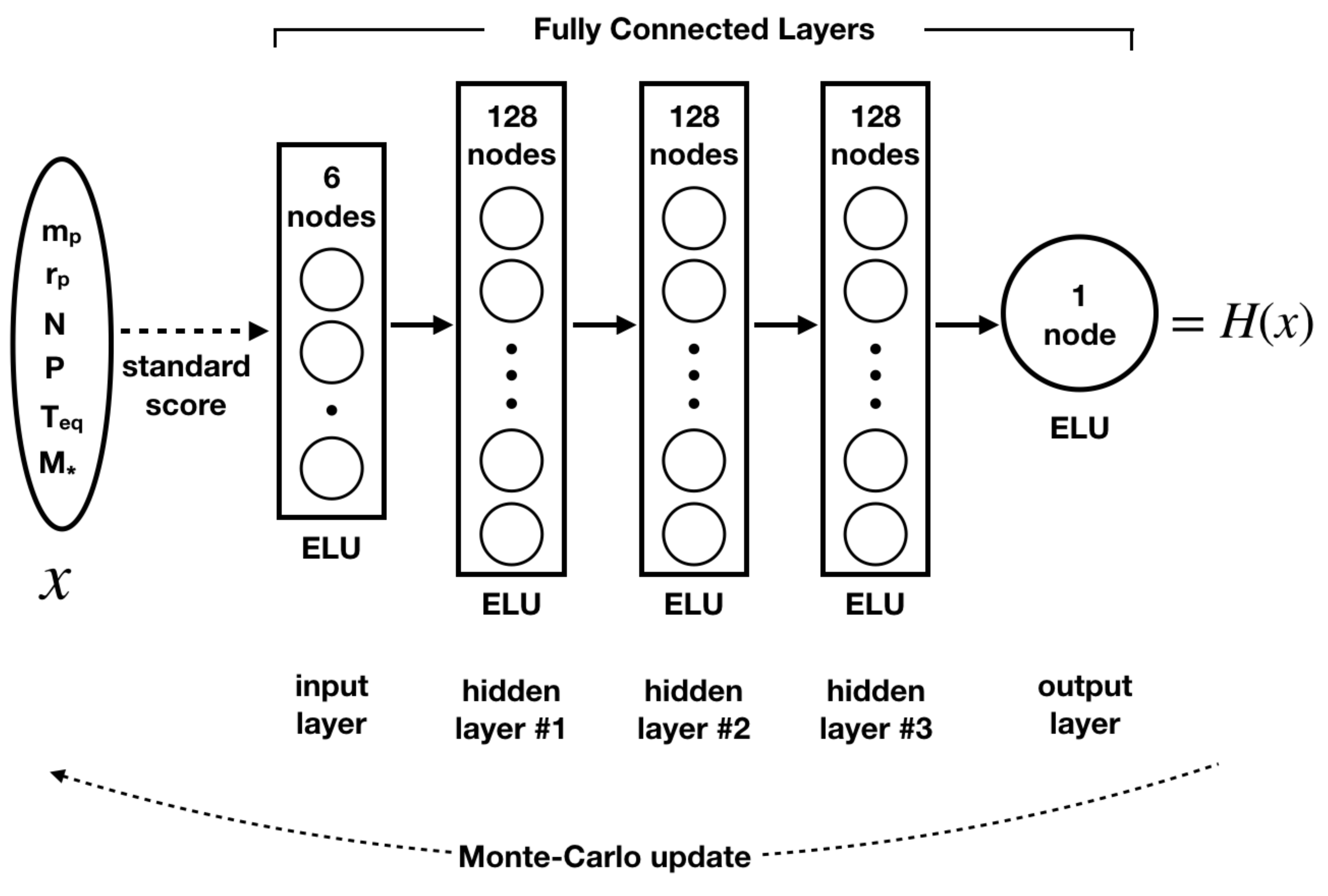}
    \caption{Architecture of the neural network. The input layer consists of the standard score (fractional number of standard deviations from the mean value) for the six planet properties, mass ($m_p$), radius ($r_p$), number of known planets in the system ($N$), orbital period ($P$), planet equilibrium temperature, ($T_{\rm eff}$) and stellar mass ($M_*$). This is passed through three hidden layers of 128 nodes to create the 1 node output as the likelihood function, $H(x)$. The likelihood function is then repeatedly sampled by Monte-Carlo to create the probability distribution for an unknown planet property.}
    \label{fig:arch}
\end{figure}

\begin{figure}
    \centering
    \includegraphics[width=\columnwidth]{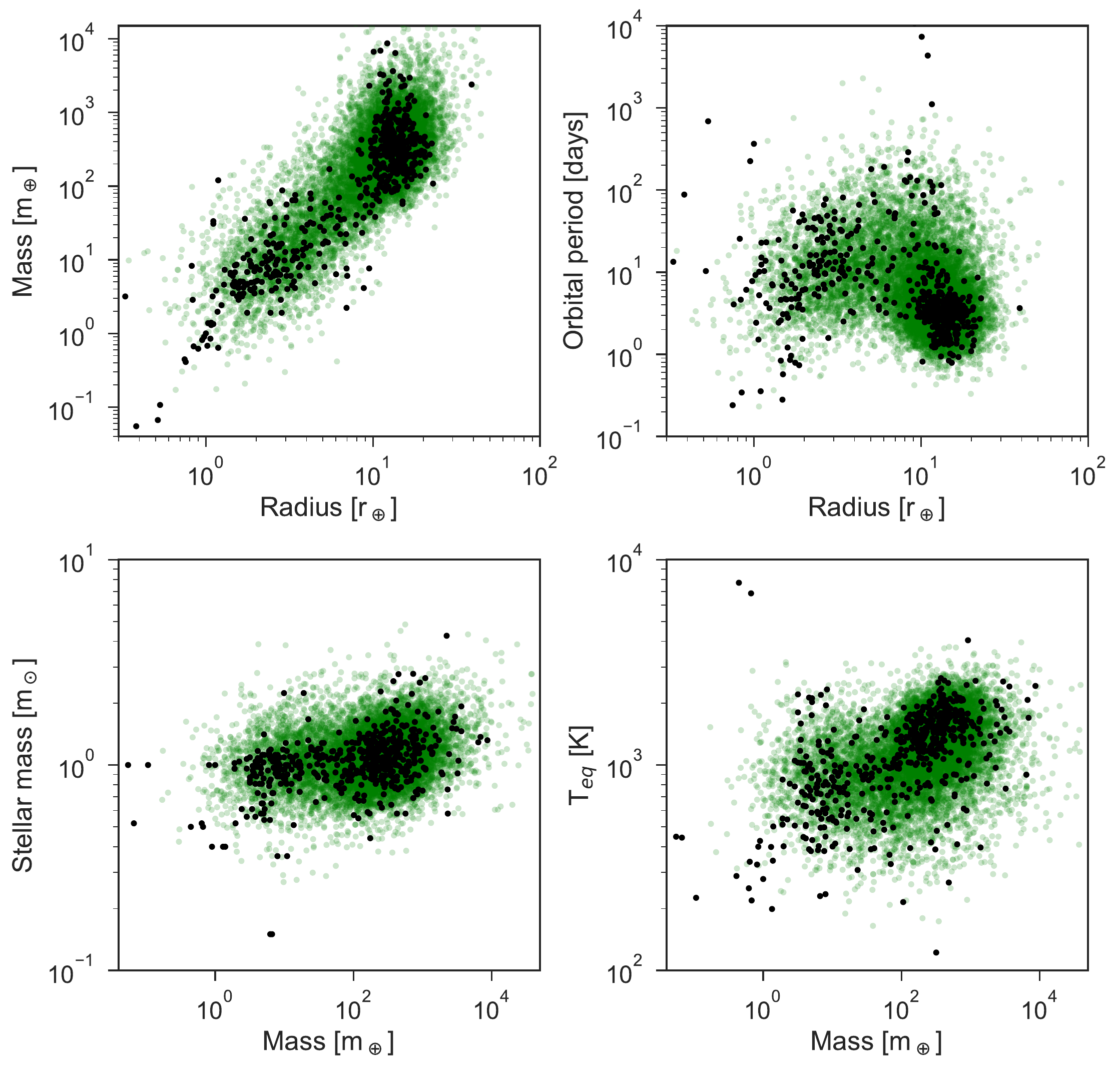}
    \caption{The density of planets inferred by the neural network. Black dots mark the measured values of confirmed planets with which the network is trained. Green shows the distribution of 15,000 generated planets whose properties have been created by the network.}
    \label{fig:altplanets}
\end{figure}

\begin{figure*}
    \centering
    \includegraphics[width=\textwidth]{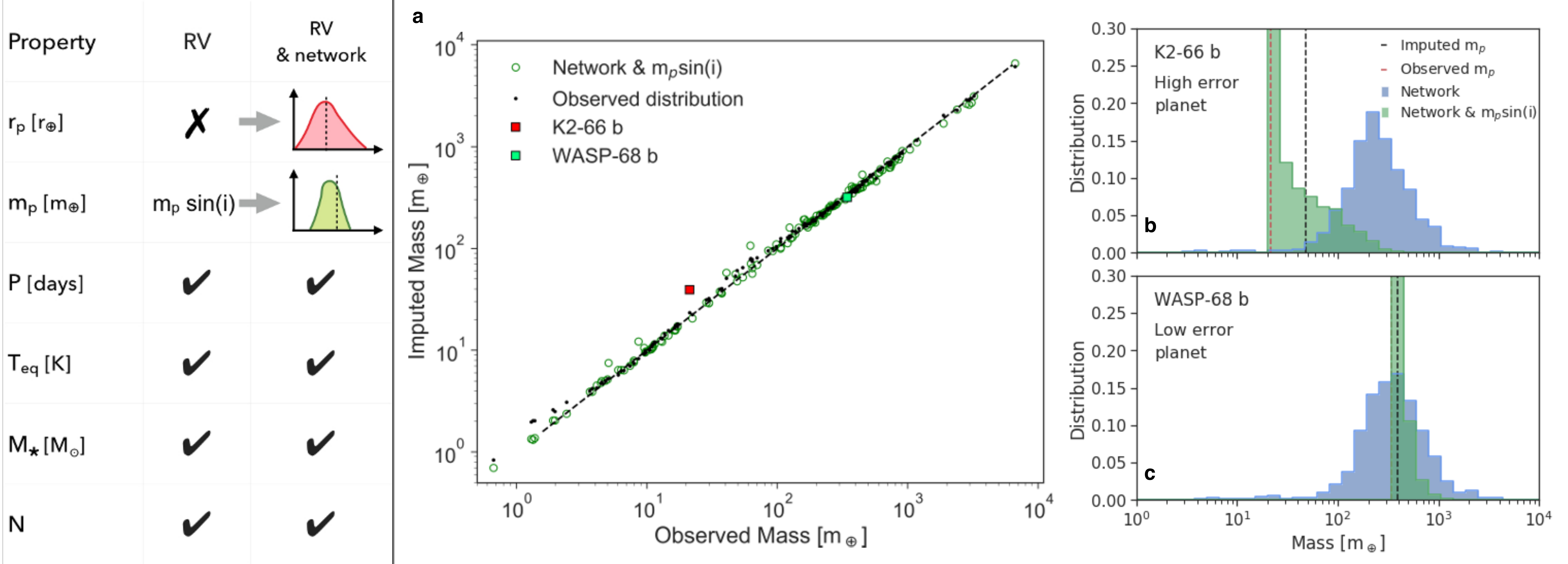}
    \caption{Test results: imputed mass values for the planets observed with the radial velocity technique. Here, the neural network is given only orbital period, planet equilibrium temperature, number of known planets in the system and stellar mass. However, for radial velocity planets the minimum mass ${\rm m_p\sin(i)}$ is known, so we weight the possible mass distribution predicted by the network (blue shaded histogram) by the probability of measuring ${\rm m_p\sin(i)}$ given the estimated mass $\rm m_p$ (see section~\ref{sec:methods}). {\bf a.} Test set data comparing imputed mass with known mass of the planet. {\bf b-c.} Mass probability distribution for a planet with a high (resp. low) error from the test data set. As the true mass is known for those planets, we chose an arbitrary inclination value to generate a ${\rm m_p\sin(i)}$ for those plots ($i=90$, edge-on observation). The vertical black dashed line is the imputed mass for the planet, defined by the mean value for the predicted distribution.}
    \label{fig:rv}
\end{figure*}

Training is performed by minimising a loss function that is the difference between the expectation value of the log probability function of the observed planets, E[$H(x)$] = $\left<{H(x)}\right>$, and the expectation value of the log probability function of the generated data points, L = E[$H(x)$] - E[$H(\tilde{x})$]. Once trained, a distribution of values for a planet property can be created by repeated sampling of the log probability function. Every distribution shown in this paper is obtained by drawing 10,000 samples from the log probability function. The quoted mass is then the average value of the distribution for masses derived from radial velocity measurements, and the modal value for the more symmetric distributions derived from transit data (see section~\ref{sec:knownmass}). The accuracy of the result is tested on a further 150 planets with the six properties that were not included in the training.

A diagram of the architecture of the neural network is shown in Figure~\ref{fig:arch}. The network itself consists of three hidden layers of size 128, with a final output layer of size 1. This takes the 6 dimensional input $\vec{x}$ of planet properties to a scalar output. Each layer has an activation function whose output, $z$, depends on the weighted sum of the 256 layer nodes. The activation function on each layer but the last is the Exponential Linear Unit (ELU) $= \exp(z)- 1$, which is a smooth non-saturating activation from $[-1,\infty]$, similar to the Rectified Linear Unit but without the discontinuity at zero. The final layer has an activation 1+ELU($z$), providing a strictly positive energy function (this prevents the energy from being unbounded below, which would cause the Monte Carlo process to fail to converge). The network is trained using the stochastic gradient descent optimisation algorithm ADAM \citep{ADAM}, which sets the step size for adjusting the weights for each layer with a learning rate $5\times 10^{-4}$. 

While we use the network in the following sections to impute the distribution of just one or two missing properties for an observed planet whose other properties are known, the density of planets within the six dimensional space inferred by the network can be explored by generating a dataset of artificial planets with all six properties imputed. This is shown in Figure~\ref{fig:altplanets}, where green points are for 15,000 generated planets, overlaid with black points of the observed dataset. As the full six dimensional space cannot be represented on a single plot, the four panels show the different properties plotted against one another. In all cases, the density of the generated planets follows that of the observed dataset closely, with outliers often reflected in a small number of green points. Two size regimes of planets are suggested in the data, consisting of worlds similar to Jupiter in mass and radius and those of the smaller super Earths, with atmospheres not large enough to have undergone a runaway collapse. This smaller sized group shows more scatter in their properties, potentially suggesting these worlds are more susceptible to environmental factors such as atmospheric stripping. While the two groups are distinguishable by the density of points, planets do populate the bridge between them.

The direct output from the network is the relative likelihood of one or more unknown feature (e.g. mass and/or radius) given information about the other known features. This can be sampled to generate a probability distribution for that feature. However, it cannot take advantage of additional information, such as the minimum mass measurement, m$_s$, that is available from radial velocity detections. As ${\rm m_s = m_p\sin(i)}$, this provides an independent probability distribution of masses, $m$, based on the possible planet orbit inclination angles, $i$. When imputing the mass for a radial velocity measurement, we therefore associate a correcting factor g(m) with the network probability distribution, p$_n$(m), that considers the fact that the inclination angle $i$ follows an sinusoidal distribution of orientations. If the random variable $\bf{I}$ has a probability density f$_I$(i) = $\sin(i)$, we can define a new random variable Z = m$_s$/sin($\bf{I}$), which then has a probability density f$_Z$(z) = $z^2$/$m\sqrt{1-z^2}$, where $z$ = m$_s$/m. Finally, we can write p(m) = p$_n$(m) f$_Z$(m$_m$/m). In Figure~\ref{fig:rv} and ~\ref{fig:mr_hab}, p$_n$ is always depicted as a blue histogram while p(m) (after including the minimum mass) is depicted as the green histogram. This is an example of an additional layer that can be applied to the network base to include information not in the observational output alone.

The training data and the number of variables can easily be expanded as future observations allow. The dependence of the network code on model parameters is discussed in the appendix, while the code and a table of the full training set are available online.

\section{Network performance}
\label{sec:knownmass}

\subsection{Mass prediction from RV observations}
\label{sec:testrv}

\begin{figure*}
    \centering
    \includegraphics[width=\textwidth]{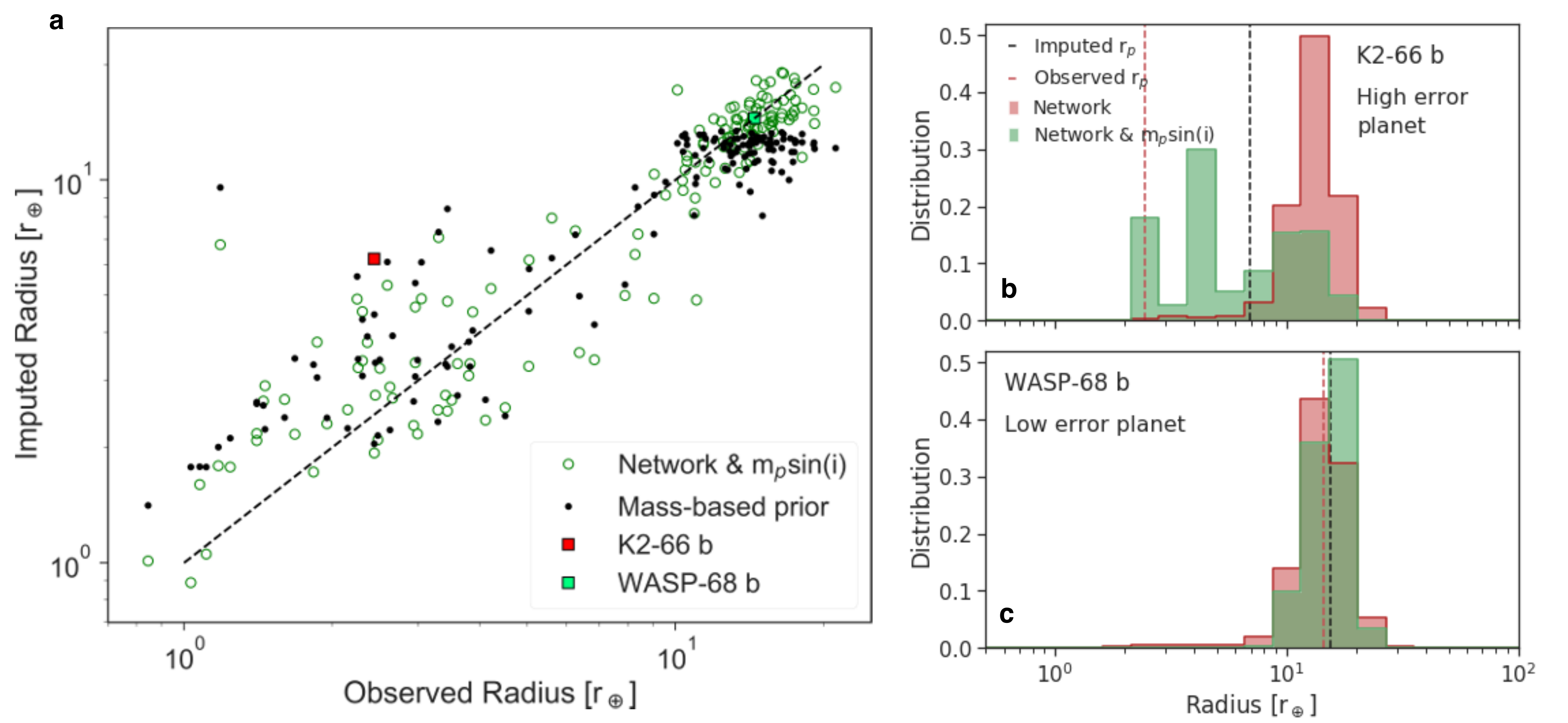}
    \caption{Test results: imputed radius values for the planets observed with the radial velocity technique. {\bf a.} Test set data comparing the imputed radius with the known radius of the planet plotted with green circles. Black dots show the results based only on an ${\rm m_p\sin(i)}$ measurement. {\bf b-c.} Radius probability distributions for the same planets as in Figure~\ref{fig:rv} with a high error (top) and low error. The distribution straight from the network is shown as a red shaded histogram, while the corrected distribution from the known ${\rm m_p\sin(i)}$ is shaded green. Vertical black dashed line is the imputed radius for the planet defined by the mean value of the corrected probability distribution, while the red dashed line is the observed radius.}
    \label{fig:rvradius}
\end{figure*}

Using our test data set of 150 planets described in section~\ref{sec:methods}, we assessed the network's ability to correctly estimate the planet mass, ${\rm m_p}$. We first treated the planets as if they had been discovered via the radial velocity technique. Such planets would have no radius measurement and only a measurement of the minimum mass, ${\rm m_p\sin(i)}$, where $i$ is the unknown orbital inclination angle. The network is therefore passed values for stellar mass, number of planets in the system, orbital period and equilibrium temperature and imputes missing values for both planetary radius and mass. The resulting mass distribution for each planet is then adjusted to account for the fact that the minimum mass ${\rm m_p\sin(i)}$ is known. Since the test planets have true mass measurements, for this procedure we create an ${\rm m_p\sin(i)}$ value for each planet based on a randomly selected angular momentum axis orientation for the orbit of the planet and line-of-sight from the telescope to give an orbital inclination angle between  $0 - 90^\circ$. This process is repeated 100 times for different inclination angles. The mass estimate is taken as the mean of the mass distribution, with the final value for the planet being the average of the mean for each inclination permutation. 

Figure~\ref{fig:rv}a shows the imputed mass plotted against the observed mass. An exact match follows the black dashed line. Green circles indicate the prediction from the network, including the correction for the known ${\rm m_p\sin(i)}$ value. As a benchmark, we compare this to predictions obtained by simply using the mass distribution of known planets instead of that from the neural network (shown as black dots), excluding the planets present in the test data. Since the distribution is approximately Gaussian in log-space, we calculate the error for each planet with a randomly selected orbital inclination angle as $\rm \epsilon = \ln({m_{p, o}/m_{p, i}})$, where m$_{\rm p,o}$ is the observed planet mass and m$_{\rm p,i}$ is the imputed value from the network. These values are averaged as the root mean square over all planets at all inclination angles to give an average error in the log of $\epsilon_{\rm net} = 0.39$ for the network and $\epsilon_{\rm obs} = 0.41$ for the benchmark using the mass distribution of known planets, demonstrating an improvement by the network on simply using the known mass distribution. The network therefore finds a mass for the planet that is on average a factor of 1.48 from the observed mass.

Although the match with the measured mass is good for the network, there are outlying points. These typically occur when the training set does not contain enough data points in that region of the parameter space. While this can improve if more data is accumulated, the accuracy of the imputed value can be gauged from the distribution of masses returned by the neural network for a given planet. An example of an entry with a high error is shown in Figure~\ref{fig:rv}b, compared with that of a low error in \ref{fig:rv}c. The shaded blue distribution in both plots shows the network mass distribution prior to the minimum mass correction. The result of applying the minimum mass adjustment is shown with the green shaded region for one case where the inclination angle was $90^\circ$ (for the point in Figure~\ref{fig:rv}a, 100 possible inclination angles are considered). The high and low error planets are also marked on Figure~\ref{fig:rv}a as red and green squares, respectively. 

The low error planet is WASP-68b, a typical inflated hot Jupiter with $\rm m_p = 1.1$\,M${\rm{_{Jup}}}$ and an orbital period of 5.1 days \citep{Delrez2014}. Its measured mass (marked by the red dashed line) is closely predicted by the network (black dashed line) at $\rm m_{p,i} = 1.2$\,M${\rm{_{Jup}}}$. The blue network and green corrected distributions strongly overlap, reflecting the fact that the measured minimum mass value sits within the high probability region of the network mass distribution. This would be true for inclinations up to about $25^\circ$ ($\rm m_p\sin(i) \simeq 0.47$\,M$_{\rm Jup} = 150$\,M$_\oplus$).

The high error planet is a more unusual case. This is K2-66b; an extremely hot sub-Neptune planet with a mass of $\rm m_p = 21.3$\,M$_\oplus$ \citep{sinukoff2017}. The planet sits in the so-called {\it photoevaporation desert} where a dearth of planets is observed with sizes between $2 - 4$\,R$_\oplus$ due to atmospheric stripping from the high levels of irradiation \citep{lopez2013}. The network (blue distribution) initially presumes this is also a hot Jupiter, but the measured minimum mass lies far outside the network's expectation. This causes the corrected distribution to take on a skewed, bimodal shape as it attempts to reconcile the two results. The imputed mass ends up sitting between the peak network value and the minimum mass, with a value of $\rm m_{p,i} = 47.5$\,M$_\oplus$. A measured minimum mass value within the tail of the network distribution means that the network believes there is a low chance of that value being close to the planet's true mass. While this cannot indicate all sources of error, the situation is a flag to alert network users that this planet is different from most discoveries and that the imputed mass could have a high error. Such a result is therefore worthy of extra consideration. 

\subsection{Radius prediction from RV observations}

\begin{figure}
    \centering
    \includegraphics[width=\columnwidth]{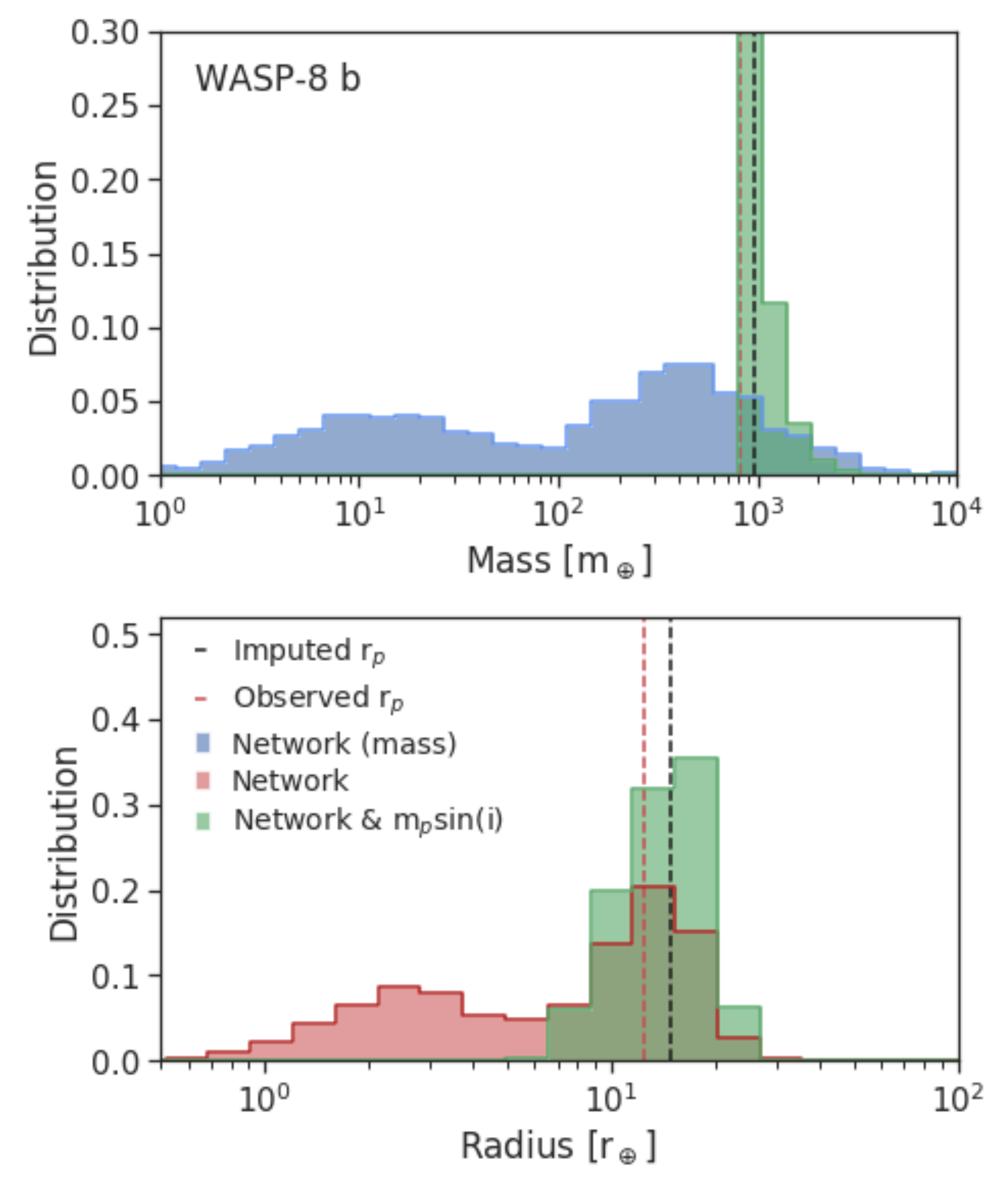}
    \caption{Imputed probability distributions for mass (top) and radius (bottom) for WASP-8b, a planet in the test set. The mass distribution returned by the network is shaded blue and the radius distribution shaded red. The green histogram show the probability distributions after being corrected with a known ${\rm m_p\sin(i)}$ value. As with the previous examples of individual planets in Figures~\ref{fig:rv} and \ref{fig:rvradius}, an inclination angle of $i = 90^\circ$ was used to generate these distributions as the true mass is known.}
    \label{fig:wasp8}
\end{figure}

In addition to mass, the neural network also predicted the radius distribution for each of the test planets presented as radial velocity data. As described in section~\ref{sec:testrv}, the network was passed values for stellar mass, number of planets in the system, orbital period and equilibrium temperature and imputed probability distributions for the unknown mass and radii values. The mass distribution was then corrected using a second distribution of possible masses from a known ${\rm m_p\sin(i)}$ minimum mass value. While the radial velocity measurement provides no minimum guide to the planet radius, each planet radius imputed by the network corresponds to an imputed planet mass. The same correction factor based on ${\rm m_p\sin(i)}$ could therefore be applied to the radius distribution. The radius for each planet was calculated in the same way as the mass, by averaging the mean of the corrected radii distribution over 100 possible orbital inclination angles for each planet.

The results for the radius prediction for the 150 test planets are shown in Figure~\ref{fig:rvradius}a as green circles. Measured as in the previous sections, the average error in the log for the neural network is $\epsilon_{\rm net} = 0.37$ or a factor of 1.4 of the observed radius. Despite the higher scatter in Figure~\ref{fig:rvradius}a compared to Figure~\ref{fig:rv}a, this is a similar value as for error on the mass due to the range of planet radii in the test data being over two orders of magnitude less than that of planet mass. Such a difference is not surprising as a comparison between the Earth and Jupiter in our Solar System shows radius of the gas giant is approximately 10.97\,R$_\oplus$, but the mass is over an order of magnitude higher at 317.8\,M$_\oplus$. 

\begin{figure*}
    \centering
    \includegraphics[width=\textwidth]{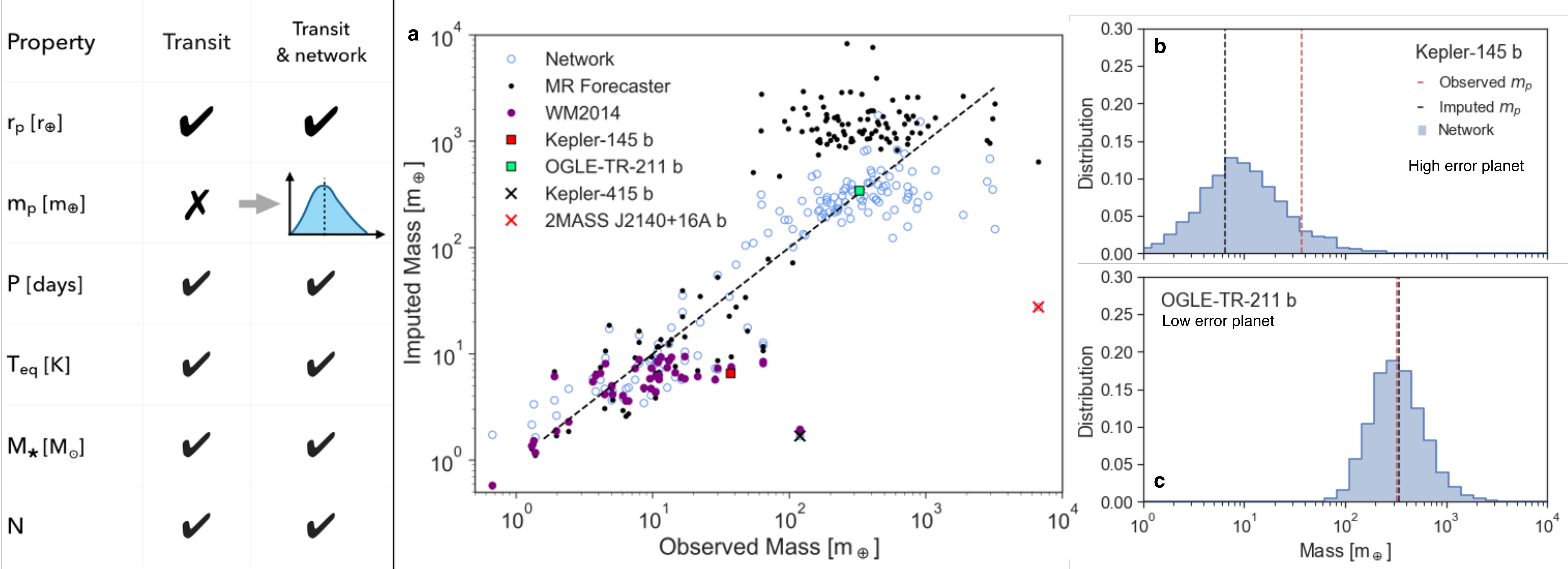}
    \caption{Test results: imputed mass values for the planets observed with the transit technique. The neural network is given planet radius, orbital period, planet equilibrium temperature, number of known planets in the system and stellar mass. {\bf a.} Test set data comparing imputed mass with known mass of the planet. {\bf b-c.} Mass probability distribution for a planet with a high (resp. low) error from the test data set (blue shaded histogram). The vertical black dashed line is the imputed mass for the planet, defined by the mode of the predicted distribution.}
    \label{fig:transit}
\end{figure*}

Creating a comparison to the neural network result is more challenging, since there is no known relationship between the measured properties in a radial velocity detection and the planet radius. We therefore created a histogram of the logarithmically average observed planet radii ($= \exp{[\left<\ln{r_p}\right>]}$) in logarithmic bins of ${\rm m_p\sin(i)}$, calculated from 100 orbital inclination angles for each planet in the training dataset. An imputed radius could then be read from the histogram based on a measured ${\rm m_p\sin(i)}$ value. This is comparable to a network trained only on mass and radius, where the histogram is replacing the gradient descent based on weights found by the network. The analysis was done in log space to minimize the effect of outliers. The results are shown as black dots in Figure~\ref{fig:rvradius}a, with an average error measurement in the log of $\epsilon_{\rm prior} = 0.49$. The relatively sharp cut-off in the distribution at large radii is due to the planet KELT-1b, whose high mass of 27.23\,M$_{\rm Jup}$ but moderate radius of 1.11\,R$_{\rm Jup}$ likely makes it a brown dwarf, rather than a gas giant \citep{Siverd2012}. The mass of this planet is spread through the top ${\rm m_p\sin(i)}$ bins and causes the radius of these bins to be underestimated for a planetary interior.

The radius probability distributions for the same two planets shown in Figure~\ref{fig:rv}b and c,  K2-66b and WASP-68b, are shown in Figure~\ref{fig:rvradius}b and c. Their values are also marked on Figure~\ref{fig:rvradius}a. The errors for the planets' imputed radii correlate with that for their masses, with K2-66b having a particularly high error while the error on the imputed radius for WASP-68b remains small. K2-66b has an imputed radius of 7.0\,R$_\oplus$, compared with the much smaller measured value of 2.4\,R$_\oplus$, consistent with the network believing this should be a more massive but inflated world. As with the mass distribution, the correction from the ${\rm m_p\sin(i)}$ distribution pushes the radius distribution into a bimodal shape as the disparate results are combined. The typical hot Jupiter WASP-68b, has an imputed radius of 15.5\,R$_\oplus$ and close measured radius of 14.5\,R$_\oplus$. The probability distributions from the network alone and the distribution after the additional ${\rm m_p\sin(i)}$ correction are almost the exact same shape.

Figure~\ref{fig:wasp8} shows the probability distributions for one other planet in the training set, WASP-8b \citep{Bonomo2017, Stassun2017}. With a measured mass of 2.54\,M$_{\rm Jup}$ and radius 1.13\,R$_{\rm Jup}$, WASP-8b is a super-sized hot Jupiter with an orbital period of 8 days. The probability distribution returned by the network is bimodal. Unlike for K2-66b, the two peaks are not caused by the incompatible probability distributions from the network and ${\rm m_p\sin(i)}$ measurement. Instead, there are two very similar mass and radius options for the planet that are statistically consistent with the planet's orbital period, host star mass, equilibrium temperature and system size. The smaller peak is a super-Earth planet about 10\,M$_\oplus$ in mass and 2.5\,R$_\oplus$ in radii. The larger peak is the hot Jupiter sized world with a mass of about 500\,M$_\oplus$ (1.6\,M$_{\rm Jup}$) and radius of 15\,R$_\oplus$ (1.4\,R$_{\rm Jup}$). This dichotomy reflects the dual populations in mass seen in Figure~\ref{fig:trainingdata} and Figure~\ref{fig:altplanets}, which both seem to overlap in the region of the network density population where WASP-8b sits. Such an example for WASP-68b demonstrates how the information available in the probability distribution returned by the network can help understand trends in the multidimensional archive data. In the case of WASP-8b, the ${\rm m_p\sin(i)}$ correction pushes the imputed value to the larger planet peak, resulting in a predicted mass of 1.35\,R$_{\rm Jup}$ and 3.0\,M$_{\rm Jup}$. 

\subsection{Mass prediction from transit observations}

The same test data can also be presented to the network as transit observations. The planet properties now include radius, but there is no guidance on the mass. The network therefore returns a distribution of planet masses based on five properties and there is no additional step to weight the distribution. The results are shown in Figure~\ref{fig:transit}. Without the weighting with a minimum mass value, the logged mass distributions are nearly symmetrical and we therefore take the peak value for the imputed mass. Figure~\ref{fig:transit}a is the equivalent plot to Figure~\ref{fig:rv}a, comparing the imputed and observed mass values. The network result (blue circles) is compared with two other one-dimensional methods for estimating planet mass based on radius alone. The black dots are the results from MR Forecaster; a probabilistic mass-radius predictor that extends from rocky planets to stars \citep{mrforecaster}. The purple dots show the empirical relation based on 65 planets smaller than 4\,R$_\oplus$ on orbits shorter than 100 days (WM2014) \citep{wm2014}.

Without the guidance from a minimum mass measurement, the scatter in Figure~\ref{fig:transit}a is significantly higher than for the radial velocity test data. Most notably, planets around a Jovian mass or higher show horizontal scatter due to the degeneracy with a stellar interior. This is strongly marked in the MR Forecaster result, likely because they only have a radius guide and also explicitly include stars in their algorithm, whereas the stellar objects included in the exoplanet catalogue are those in the grey area between planet and brown dwarf. Measured the same way as for the radial velocity test data, the average error in the log for the neural network is $\epsilon_{\rm net} = 0.98$ (a factor of 2.7 from the observed radius) while MR Forecaster is $\epsilon_{\rm MRF} = 1.6$ and the empirical WM2014 is $\epsilon_{\rm WM} = 1.02$. The network performs similarly to WM2014, but over the full range of planet masses. It offers an improvement of 1.6 over the MR Forecaster, mainly due to the improvement at large radii. If all three methods are restricted to planets smaller than 4\,R$_\oplus$, then the error in the log is $\epsilon = 1.0$ in all three cases.

Figures~\ref{fig:transit}b and \ref{fig:transit}c show examples of the mass distribution for the high and low error cases, respectively. The low error case is another hot Jupiter, OGLE-TR-211b, with a measured and network imputed mass of 1.03\,M$_{\rm Jup}$ \citep{udalski2008}. The high error planet is once again an usual case. Kepler-145b has a measured mass of 37.1\,M$_\oplus$ and radius 2.65\,R$_\oplus$, which gives an average density of $\rho = 11.0$\,g/cm$^3$ and places the planet among the largest rocky planets known to-date \citep{xie2014}. The network underestimates the mass, predicting a value of $6.6$\,M$_\oplus$ and giving a density of $\rho_{\rm i} = 1.9$\,g/cm$^3$, consistent with a planet with a thicker, Neptune-like atmosphere. This prediction is agreement with the majority of planets currently discovered, which seem to accrete thick atmospheres once their radius reaches $1.6$\,R$_\oplus$ \citep{rogers2015}.

Two other points are marked with an `X' on Figure~\ref{fig:transit}a as examples of planets that are extreme outliers. 2MASS J2140+16A is a Jupiter-sized planet with a measured mass of  20.95\,M$_{\rm Jup}$ and is described in its discovery paper as a brown dwarf in a low mass binary star system \citep{konopacky2010}. The long 20 year orbit is exceeded by only three other entries in the training and test data set: Saturn, Uranus and Neptune. The lack of comparative data to learn from and the stellar nature of the object resulted in the network significantly underestimating the planet mass as $0.09$\,M$_{\rm Jup}$. 

The second extreme outlier is Kepler-415b, a planet with a recorded measured mass of 120\,M$_{\oplus}$ and radius 1.2\,R$_{\oplus}$, giving an unphysical average density of $\rho = 373.1$\,g/cm$^3$\,\citep{hadden2014}. The planet was discovered through transit timing variations and has an extremely large error bar of the same order as its mass. The true mass value is therefore likely much smaller and the neural network suggests it should be about $1.7$\,M$_{\oplus}$.

\section{The masses of small temperate planets}
\label{sec:smallplanets}

\begin{figure}
    \includegraphics[width=\columnwidth]{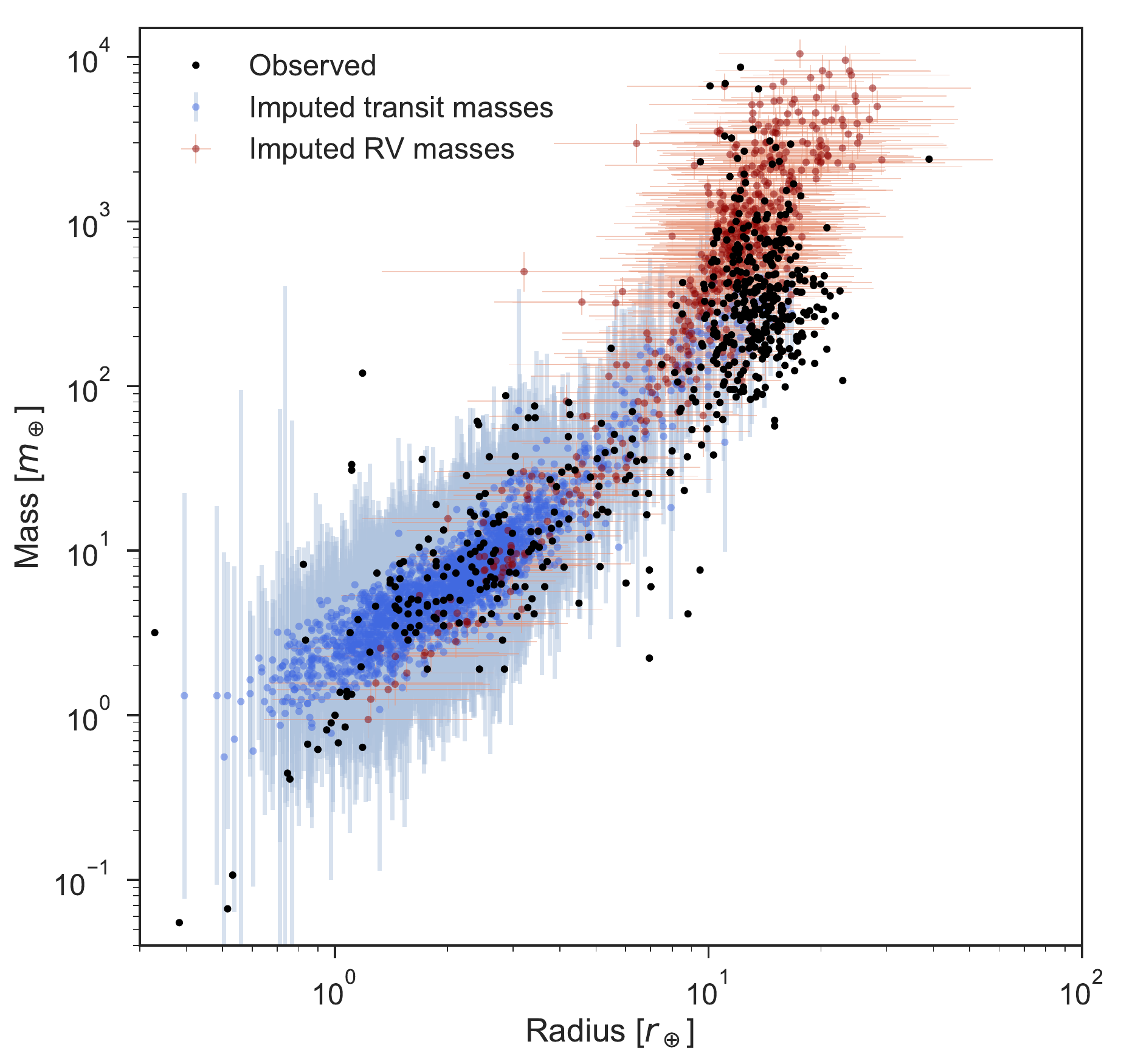}
    \caption{Neural network predictions for the mass (and in the case of the radial velocity observed planets, also radius) for planets observed with either radial velocity or the transit technique so without a measured mass (and radius) value. Error bars indicate the width of the probability distribution.}
    \label{fig:mr_all}
\end{figure}

\begin{figure*}
    \centering
    \includegraphics[width=\textwidth]{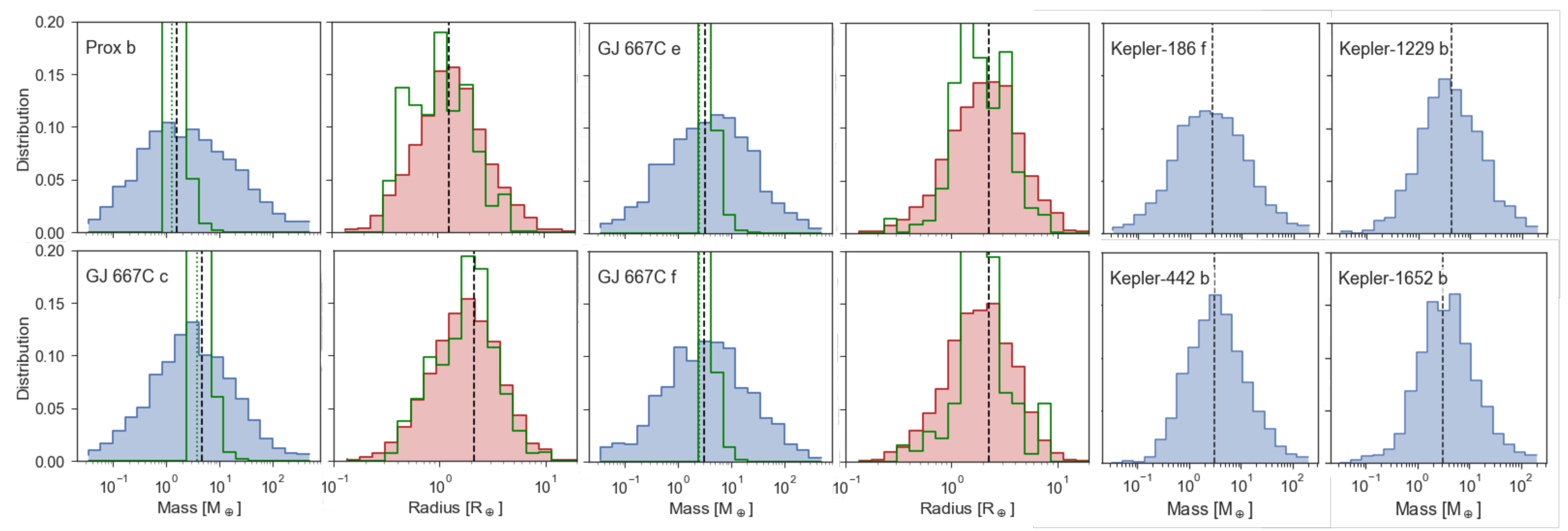}
    \caption{Imputed mass and radii probability distributions values for six 
    likely rocky (based on either their radius or minimum mass) planets that spend more than 90\% of their orbit within the habitable zone. The vertical black dashed line shows the imputed value from the distribution. In the case of the planets discovered with the radial velocity method, the green histogram shows the corrected distribution after the application of the measured minimum mass.}
    \label{fig:mr_hab}
\end{figure*}

Using the neural network, we imputed the expected mass values for all confirmed planets listed in the exoplanet archive for which there is no mass measurement. For those entries discovered using the radial velocity technique, we also imputed the expected radius. The resulting mass-radius plot for these worlds is shown in Figure~\ref{fig:mr_all}. Planets detected with the transit technique are shown in blue with their imputed mass measurement, while the radial velocity planets are in red with imputed mass and radii. The error bars shown for each planet are derived from the width of the mass and radii distributions measured via the root mean square deviation, $\rm \sigma = \sqrt{\sum_n^N \left(\ln{m_{p,i}} - \ln{m_n}\right)^2/N}$, where $\rm m_{p,i}$ is the imputed mass and $\rm m_n$ are the N masses in the generated distribution. The upper and lower bounds then become ${\rm \exp(\ln(m_p) + \sigma) - m_p}$ and ${\rm m_p - \exp(\ln(m_p) - \sigma)}$ respectively, and similarly for planet radius. There is no true error measurement on the imputed planet properties, but these limits indicate how peaked the probability distribution is about the imputed value. Smaller error bars indicate a narrower probability distribution and therefore a smaller range of masses (or radii) that are consistent with the statistical distribution of planets discovered by the neural network. Black dots show the planets for which both mass and radius have been measured. The network overlays the observed values well, with the widest probability distributions for the very smallest planets, where the parameter space (marked out by the black dots) is particularly sparse.

\begin{table}
%\newpage
%\begin{longtable}{lcccc}
\centering
\caption{Imputed mass and radii for planets that spend at least part of their orbit within the optimistic habitable zone.}
\medskip
\begin{tabular}{p{1.6cm}cccc}
RV planets\\
\hline
 & m$_{\rm p}$sin(i) [M$_\oplus$] & m$_{\rm p}$ [M$_\oplus$] & $P_{\rm i > 80^\circ}$[\%] & r$_{\rm p}$ [R$_\oplus$]\\
\hline
     GJ 1132 c   & 2.5 & $3.3^{+0.94}_{-0.73}$   & 31.1 & $1.8^{+1.7}_{-0.86}$\\
     GJ 273 b    & 2.9 & $3.7^{+1.1}_{-0.87}$   & 15.0 & $2.3^{+1.7}_{-0.99}$ \\
     GJ 3323 c   & 2.2 & $2.9^{+0.85}_{-0.66}$  & 25.9 & $1.5^{+1.2}_{-0.69}$  \\
     GJ 667C c  & 3.8 & $4.8^{+1.4}_{-1.1}$   & 10.3 & $2.1^{+2.1}_{-1.1}$\\
     GJ 667C e  & 2.5 & $3.2^{+1.0}_{-0.78}$   & 24.3 & $2.2^{+1.9}_{-1.0}$ \\
     GJ 667C f  & 2.5 & $3.1^{+0.92}_{-0.71}$  & 36.0 & $2.3^{+1.9}_{-1.0}$ \\
     GJ 832 c    & 5.4 & $7.0^{+2.2}_{-1.7}$   & 14.4 & $2.6^{+1.7}_{-1.0}$  \\
     Prox Cen b & 1.3 & $1.6^{+0.46}_{-0.36}$ & 29.2 & $1.3^{+1.2}_{-0.62}$ \\
     Ross 128 b  & 1.3 & $1.8^{+0.56}_{-0.43}$ & 15.5 & $1.6^{+1.1}_{-0.65}$ \\
     Wolf 1061 c & 3.5 & $4.2^{+1.2}_{-0.91}$   & 26.4 & $2.1^{+1.8}_{-0.97}$  \\
\hline
\multicolumn{2}{l}{Transit planets} \\
\hline
 & r$_{\rm p}$ [R$_\oplus$] & m$_{\rm p}$ [M$_\oplus$] \\
\hline
   K2-3 d & 1.5 & $4.6^{+14.3}_{-3.5}$ \\
   K2-72 e & 1.3 & $3.5^{+16.9}_{-2.9}$\\
   Kep-1229 b & 1.4 & $4.3^{+13.4}_{-3.3}$\\
   Kep-1652 b & 1.6 & $3.1^{+8.7}_{-2.3}$\\
   Kep-186 f & 1.2 & $2.7^{+21.0}_{-2.4}$\\
   Kep-296 e & 1.5 & $2.8^{+14.4}_{-2.3}$\\
   Kep-438 b & 1.1 & $1.9^{+5.0}_{-1.4}$\\
   Kep-442 b & 1.3 & $3.1^{+11.0}_{-2.4}$\\
   Kep-452 b & 1.6 & $5.9^{+21.8}_{-4.6}$\\
\end{tabular}
\label{table:hz}
\end{table}

In Figure~\ref{fig:mr_hab}, we show the generated mass and radius distributions for planets that are likely to be rocky and spend over 90\% of their orbits within the conservative boundaries for the habitable zone \citep{hzgallary, habzone}. A radii cut-off of $1.6\,R_\oplus$, with the equivalent minimum mass for an Earth-like composition of $6\,M_\oplus$, was used to determine if the planet was likely rocky \citep{rogers2015}. The imputed values for these and for planets that spend at least part of their orbit within the conservative habitable zone limits are shown in Table~\ref{table:hz}. For the radial velocity planets, we also include the probability that the planet's orbital inclination angle is greater than 80$^\circ$. This is calculated from the distribution of inclination angles directly derived from the distribution of masses, m, shown in Figure~\ref{fig:mr_hab} via ${\rm i = arcsin(m_p\sin(i)/m)}$ for the planet's measured ${\rm m_p\sin(i)}$. Orbiting within the habitable zone, the inclination angle needed for these planets to transit exceeds 89$^\circ$. Due to the error bars on the mass estimates, this one degree range is too small to estimate probabilities meaningfully, but the fraction of possible inclinations that sit between 80$^\circ$ - 90$^\circ$ provide a suggestion of how edge-on the orbit it likely to be. This information could be utilised by missions such as CHEOPS, which aim to hunt for transits in systems previously discovered by radial velocity \citep{broeg2013}. The upper and lower bounds on the imputed values represent the root mean square deviation, $\sigma$, of the probability distribution as described above. The network values imply that the majority of planets whose minimum mass suggests a rocky composition, may in truth be either ocean worlds or mini-Neptunes. In the radial velocity group, only Proxima Centauri b and GJ 3323c have densities high enough to be considered rocky. The transit planets all have terrestrial densities, reflecting the fact that their radius measurement is a better guide to this bulk property than just the lower limit on the mass. 

\section{Alternative density models}
\label{sec:densitymodels}

\begin{figure*}
    \centering
    \includegraphics[width=\textwidth]{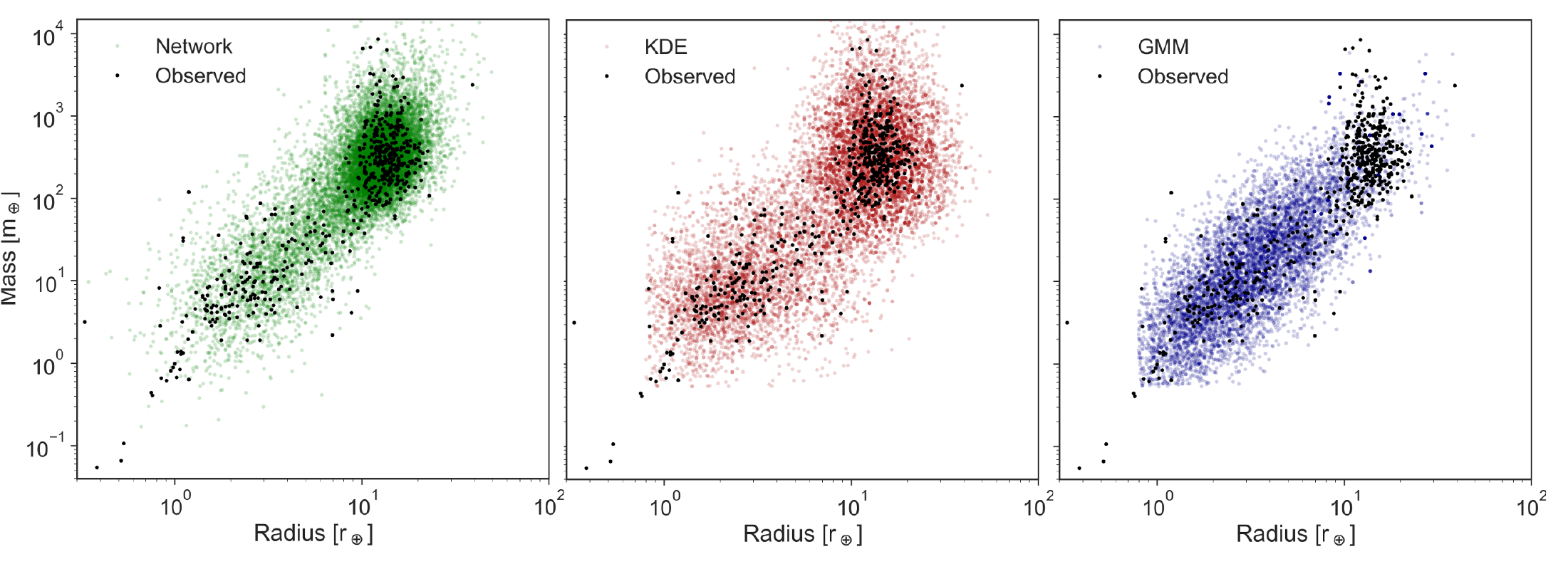}
    \caption{15,000 Generated planets from probability density spaces created by the neural network (left, green), KDE (centre, red) and GMM (right, blue). Black dots show the observed catalogue data.}
    \label{fig:distributions}
\end{figure*}

The purpose of introducing the neural network was to more fully utalise the information available in the multidimensional data of the exoplanet archive. This is not, however, the only technique that can construct and use a multidimensional probability density space to impute missing values. Two common density estimators that can be used in this way are Kernal Density Estimation (KDE) and the Gaussian Mixture Model (GMM) \citep{Silverman1986}. As with the neural network, both KDE and GMM are non-parametric methods to impute variables.  

KDE is similar to a continuous replacement for a discrete histogram in the required number of dimensions. The main parameters for the KDE method are the choice of kernal (which controls how neighbouring data points contribute to the local density) and the bandwidth, which alters the shape of the kernal to affect the smoothness of the fit. Here we used a Gaussian kernal and a bandwidth of 0.35, which produced the best results during a parameter search of bandwidth values extending from 0.05 - 1.5. For one instance of the radial velocity test data, this bandwidth failed to find a planet mass as the minimum mass (${\rm m_p\sin(i)}$ value) fell outside the imputed distribution. In that case, the routine rolled back to a bandwidth of 1.0 and recalculated. 

GMM creates a probability density by assuming the data points can be represented from a linear combination of a finite number of Gaussian distributions. The main parameter is the number of components, which controls the number of Gaussian types in the mixture. We used 2 components for the GMM, which performed the best on the test data for a parameter search between 1 and 10 components. As with the KDE model, there was one planet in the radial velocity test set whose minimum mass fell outside the GMM distribution. We likewise rolled back to a 1 component model at that time.

\begin{table}
\centering
\caption{Average error on the test data for different probability density estimators. $\epsilon_{\rm Net}$ are the results for the Boltzmann neural network used in this paper, $\epsilon_{\rm KDE}$ and $\epsilon_{\rm GMM}$ are for the KDE and GMM methods, respectively.}
\begin{tabular}{lccc}
\hline
 & $\epsilon_{\rm Net}$ & $\epsilon_{\rm KDE}$ & $\epsilon_{\rm GMM}$ \\
\hline
     RV Mass        & 0.39 & 0.40 & 0.46 \\
     RV Radius      & 0.37 & 0.39 & 0.48 \\
     Transit Mass   & 0.98 & 1.2  & 0.96  \\
\hline
\end{tabular}
\label{table:distributions}
\end{table}

The mass and radius of 15,000 generated planets with the Boltzmann neural network, KDE and GMM are shown in Figure~\ref{fig:distributions}. The left-hand  network plot is the same as the top-left panel in Figure~\ref{fig:altplanets}, repeated here for ease of comparison. The average error for each method on the test data is shown in Table~\ref{table:distributions}, where the error is calculated as in section~\ref{sec:testrv}. RV Mass and RV Radius denote results for imputing the mass and radius when the test data is presented as radial velocity observations, while Transit Mass is the test data for transit observations.

A visual inspection of the generated planet distribution in Figure~\ref{fig:distributions} shows similar shapes for the network and KDE density spaces, with the KDE showing a broader scatter in the properties. By contrast, the GMM matches the distribution of small planets reasonably well, but does not develop a distinct population group for the high mass worlds. This is reflected in the errors shown in Table~\ref{table:distributions}. KDE does slightly worse than the network across all three scores, performing comparatively worst on the transit data were there is no minimum mass guide. GMM does particularly poorly on the RV test data, which can be seen from Figure~\ref{fig:mr_all} to dominate the population of massive planets. Over the smaller transit planets, GMM marginally performs the best of all three methods. 

The comparison suggests that all three density probability methods can be used to impute values in the multidimensional dataset, with the network and KDE overall reproducing the features in the observed density more accurately than GMM.

\section{Transmission of observational biases}
\label{sec:bias}

\begin{figure}
    \centering
    \includegraphics[width=\columnwidth]{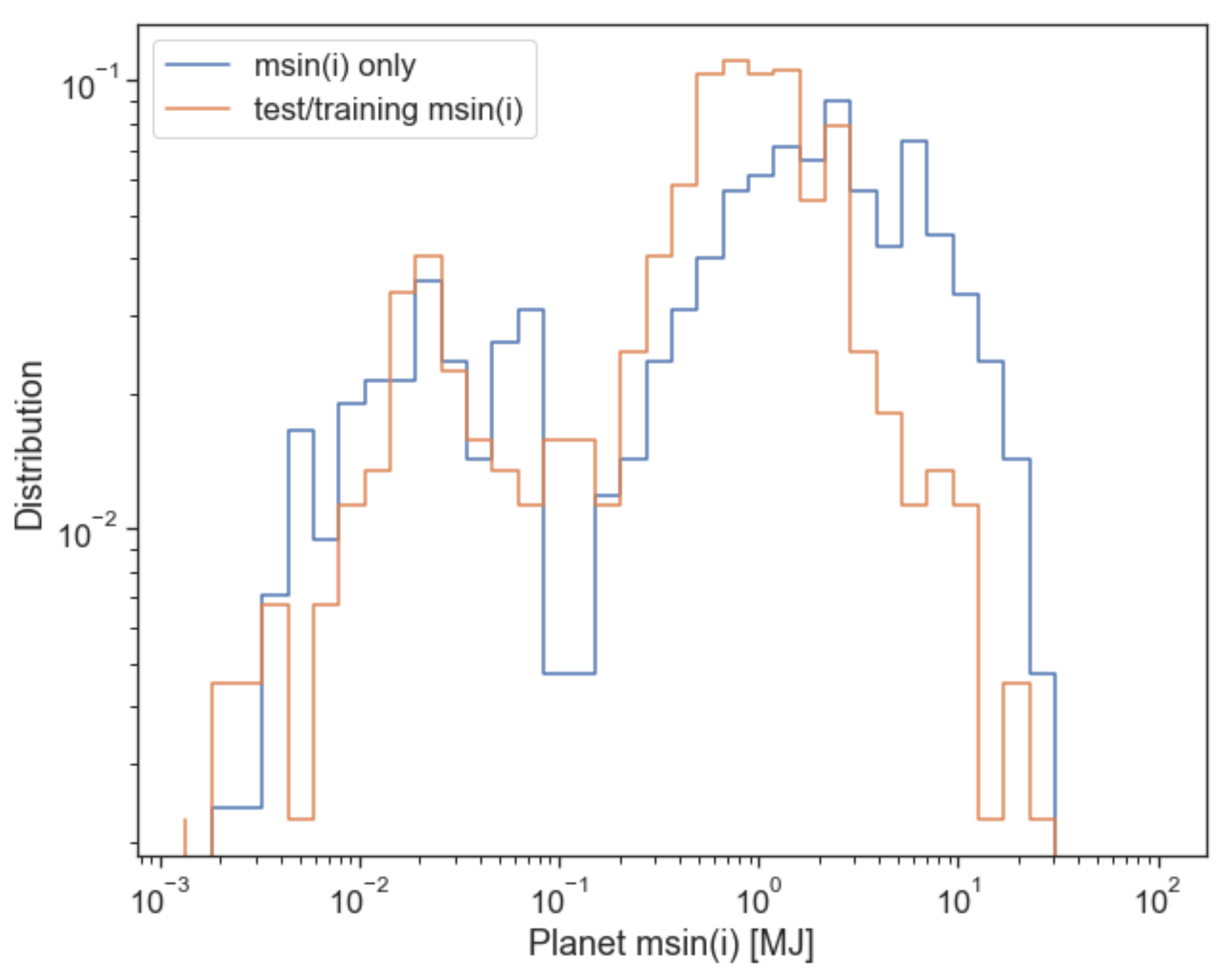}
    \includegraphics[width=\columnwidth]{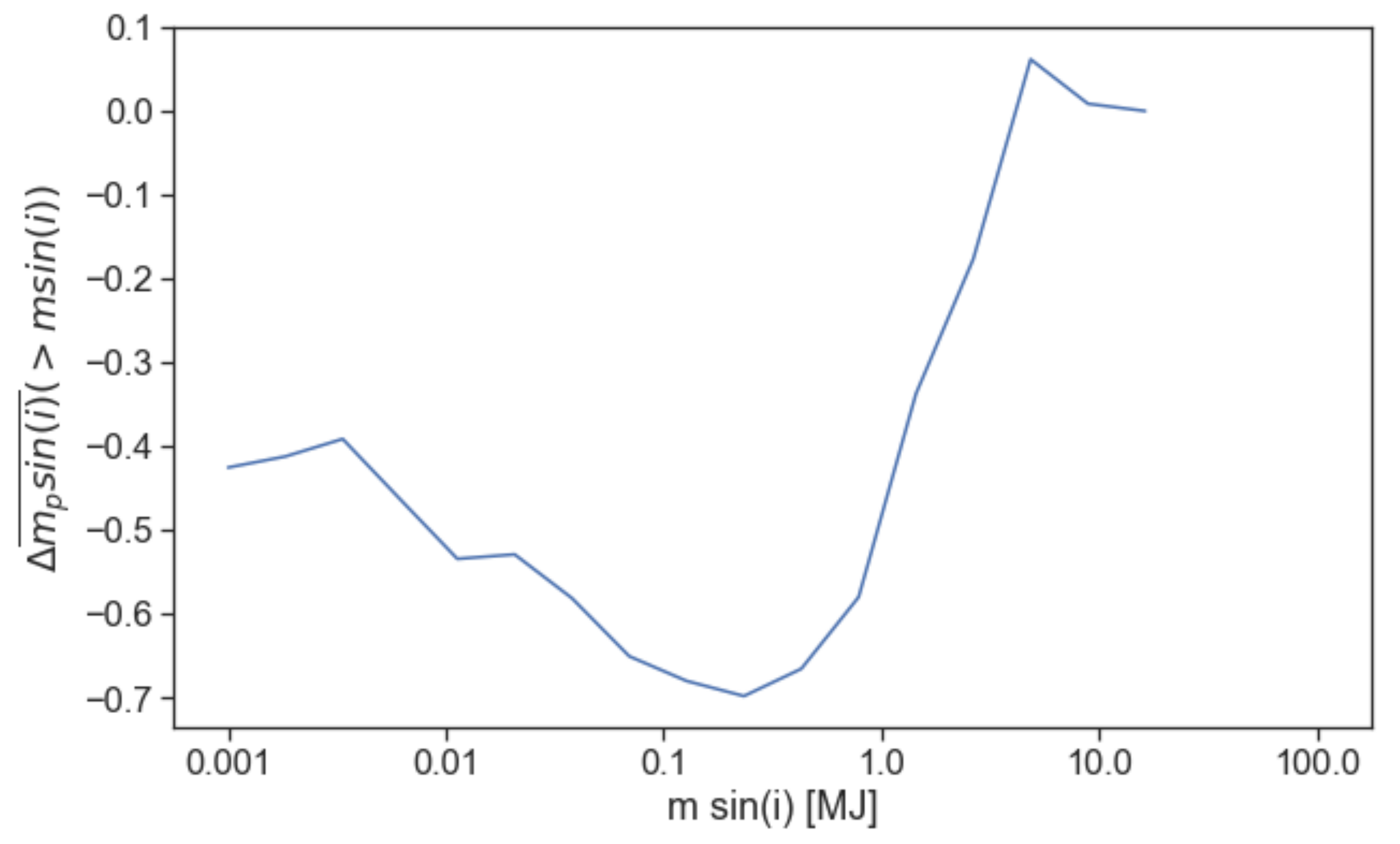}
    \caption{Top panel shows the distribution of $m\sin(i)$ values in the network training set (orange) and those of the planets whose mass was estimated in the paper. Bottom panel shows the difference in the mean $m\sin(i)$ value of the two distributions for $m_p\sin(i) > m\sin(i)$.}
    \label{fig:msini}
\end{figure}

One concern with the network training on the catalogue of measured exoplanet properties is that this dataset contains observational biases. These biases will be transmitted to the results obtained by the network, resulting in the learned probability function becoming the product of current instruments capabilities and not only the underlying planet property distributions. This can produce large errors where the training data bias and bias in the observation requiring imputed values differ. As the neural network is effectively a non-parametric interpolation technique, it cannot successfully extrapolate beyond the bounds of the data on which it has been trained. 

A particular issue for our training dataset is that the majority of entries are observed by both the radial velocity and transit techniques. The planets whose masses were imputed and plotted in Figure~\ref{fig:mr_all} and \ref{fig:mr_hab} have at most one of these measurements. This may result in these planets having different probability functions than that found by the network due to orbiting dimmer or more active stars that potentially hinder the second detection. 

Quantitative assessment of this effect is extremely difficult. However, one comparison that can be made is the distribution of measured ${\rm m_p\sin(i)}$ values present in our training set data with the distribution of ${\rm m_p\sin(i)}$ of the planets whose mass was imputed in this paper in Figure~\ref{fig:mr_all}. In this case, we are not using a set of randomly selected inclination angles, but are comparing with 444 planets in the training and test set that were detected with the radial velocity technique, providing an ${\rm m_p\sin(i)}$ value prior to the orbital inclination (and mass measurement) being found from an additional detection by the transit. As the minimum mass and true mass are related quantities, distributions for the minimum mass that differ in range and shape would suggest the training data is insufficient to predict the mass these planets. However, the range of the two distributions strongly overlap and display a similar structure as can be seen in the top panel of Figure~\ref{fig:msini}. This implies that the imputed mass values are being derived from trends learned from planets with a similar statistical distribution. The bottom figure panel shows the difference in the mean value for increasing ${\rm m_p\sin(i)}$ value, which remains below our stated network error for planets observed with the radial velocity technique. This does not suggest that such observational bias is negligible, but seems to be within the errors presented throughout this work.  

Further compensation for bias or theoretically derived adjustments to the observational dataset could be applied as corrections to the network distribution, as demonstrated for the range of inclination angles for the radial velocity planets in Figure~\ref{fig:rv}. The advantage of using the observed dataset as a prior in this manner is that it allows the maximum possible observed planetary properties to be included in the analysis, which is the ground truth of any planet formation theory as this is what has actually been measured. 

\section{Conclusions}
\label{sec:conclusions}

In this paper, we have developed a neural network to impute a probability density for planets that can be used to impute missing values in the exoplanet catalog based on the observed planetary properties. The purpose of this work is to develop a method for utilising the maximum possible information present in the observed data in analysis. 

The results presented here focus on the network's capability for estimating planet mass; a particularly challenging property to measure due to the need to employ multiple detection techniques. The network estimated masses over the four orders of magnitude present in the exoplanet archive catalogue. The average error over this range was a factor of 1.48 from the observed mass value for planets detected with the radial velocity technique, and 2.7 for those found with the transit technique, for which there is no minimum mass guide. For the estimation of planet radius from a radial velocity detection, the average error was factor of 1.4 from the observed radius.

The structure of the returned probability distribution of a property imputed by the network contains information about that property. Broad or bimodal distributions point to a wider range of values consistent with the multidimensional parameter space of currently observed planets. 

The overall performance of the network for imputing the mass is similar or exceeds previous work that imputes mass based on the planet radius alone, and can be used over the full size distribution from rocky to gas giant planets. The network can also be used to impute additional missing properties, such as planet radius. The strength of this technique is that no a priori functional dependence needs to be specified between the properties in the data, allowing for a scalable technique that can discover trends in the data beyond those of current theories. As this network returned the probability distribution for the imputed property, Gaussianity also does not need to be assumed. 

However, it should be recognised that the lack of prior assumptions about the data means that the technique will impute a value equivalent to an observed measurement. The network has no knowledge of instrumental error and so can not attempt to estimate the true planet property separately from the likely measured value. Corrections for effects such as sample bias must therefore be applied to the network result in the same fashion as an observation.

Other methods to estimate probability densities can be used in a similar way to the network presented in the paper. In particular, both KDE and GMM can be used to generate planet properties using information from multiple variables. In the tests performed here, these methods performed reasonably but did appear fit the overall distribution of planets as tightly as the network.

As the exoplanet catalog continues to expand, further property fields such as stellar composition, eccentricity and inclination can be added to this methodology once a reasonable amount of data has been acquired for training. Missions such as GAIA for stellar properties, CHEOPS and the radial velocity follow-ups for TESS observations will improve the amount of data available, allowing networks such as the one presented here to more accurately impute values where none can be acquired.

\section*{Acknowledgements}

This research has made use of the NASA Exoplanet Archive, which is operated by the California Institute of Technology, under contract with the National Aeronautics and Space Administration under the Exoplanet Exploration Program. E.J.T. was partially supported by JSPS Grant-in-Aid for Scientific Research Number 15K0514. The authors would like to thank the anonymous referee whose comments significantly improved the paper.

\appendix

\section{Network sensitivity and bias}
\label{sec:networktest}

The network training and computation properties have two main parameters defined by the user. The first is the number of Monte-Carlo steps (MCS) taken by the network during training to sample the trained likelihood function. The second is independent of training and is the number of times the resultant likelihood function is sampled to get a given property distribution (Ndist). Each of those Ndist point samples the likelihood function over the same number of steps the network was trained with (MCS). Should the number of MCS be too small, then the returned imputed value will depend on the initial, randomised, starting value for the property. If Ndist is too small, then the drawn distribution will not record the spread of possible values accurately. 

\subsubsection{Variations in number of MC steps (MSC)}
\label{sec:msc}

\begin{figure}
    \centering
    \includegraphics[width=10cm]{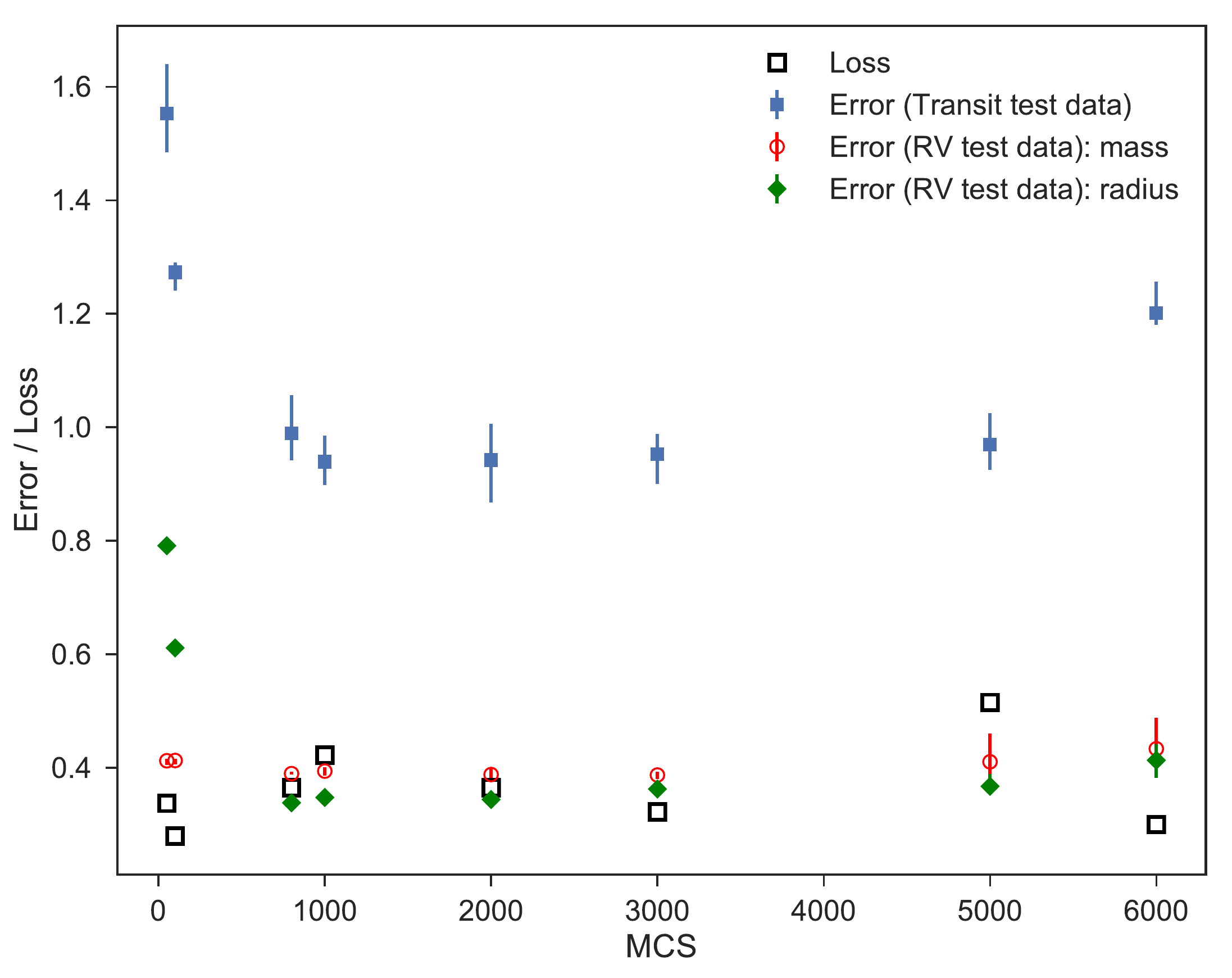}
    \caption{Plot of the average error for the test data set when represented at radial velocity data (RV test data: equivalent to Figure 1 in the main text) and transit data (Transit test data: equivalent to Figure 2 in the main text) for variations in the number of Monte Carlo steps. The radial velocity result includes the minimum mass weighting over 100 possible minimum mass values. The black squares show the average network loss value.}
    \label{fig:mcs}
\end{figure}

The variation with MCS is plotted in Figure~\ref{fig:mcs}, where Ndist = 2000 in all cases. The orange points show the error in the test data set when it is presented to the network as radial velocity data, including the weighting from the minimum mass measurement (see main text and Methods). This is equivalent to the error calculated from Figure 1a of the main text. The blue points show the error in the same test data when presented as transit data. This is equivalent to Figure 2a in the main text. The error bars shown on each point mark the range of values obtained for each of the four network outputs. Also plotted is the average network loss over the four outputs considered (see Methods). The variation error in the results flattens after about 1000 steps, decreasing slightly to a minimum at 2000 steps before starting to increase again as the network becomes over-sampled. We therefore selected MCS=2000 for the results presented in the paper.

\subsubsection{Variations in the sample size of the likelihood function, Ndist}

\begin{figure}
    \centering
    \includegraphics[width=10cm]{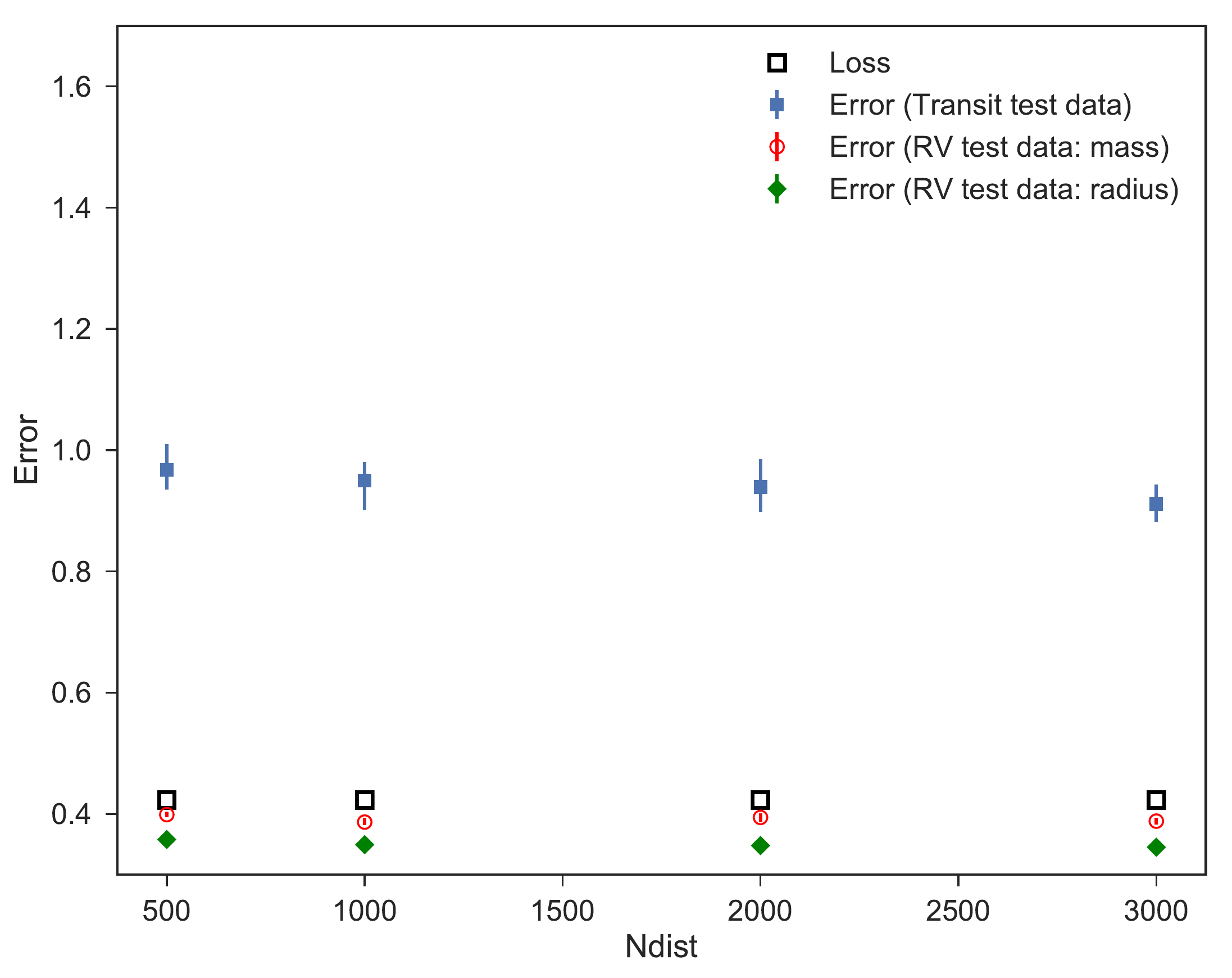}
    \caption{Plot of the average error for the test data set when represented at radial velocity data (RV test data: equivalent to Figure 1 in the main text) and transit data (Transit test data: equivalent to Figure 2 in the main text) for variations in the distribution size. The radial velocity result includes the weighting for the minimum mass values as in section~\ref{sec:msc} and the main text.}
    \label{fig:ndist}
\end{figure}

The variation with Ndist for the radial velocity and transit test data is shown in Figure~\ref{fig:ndist}, for a fixed MCS = 1000. There is no significant variation across the four values we considered, so we therefore selected Ndist = 2000 for our results. 

\subsubsection{Train/test set permutation variability}

\begin{figure}
    \centering
    \includegraphics[width=10cm]{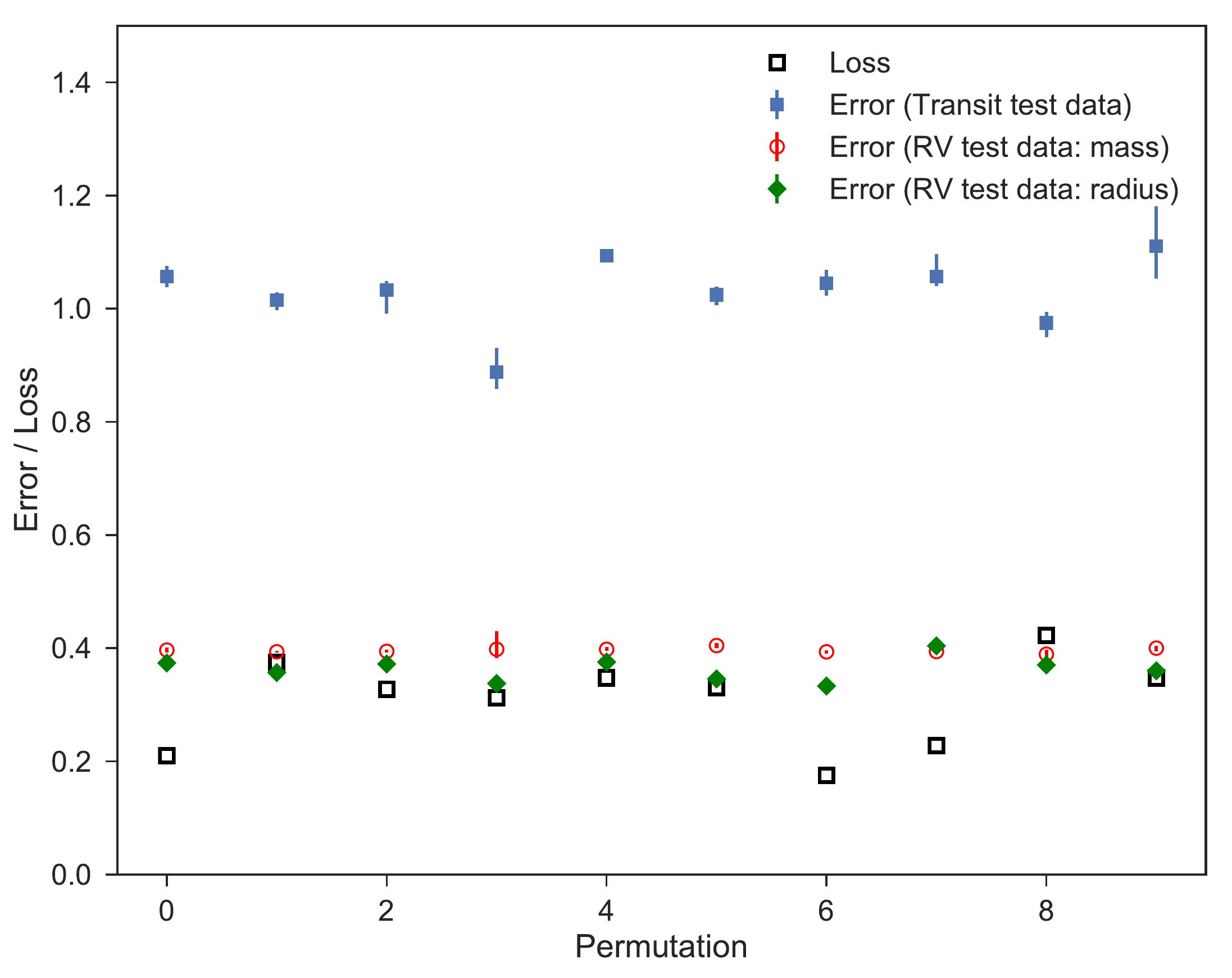}
    \caption{Plot of the average error for the test data set when represented at radial velocity data and transit data for permutations of the data split between training and test data. The network loss function is also shown as black squares.}
    \label{fig:perm}
\end{figure}

There is also potential for variation in the results due to different divisions in the planets selected for the training and test data. To ensure this has not affected our results, we train ten different networks with different random permutations. The resulting error and network loss value is shown in Figure~\ref{fig:perm}. There is no significant variation between permutations. We therefore used permutation 0 for the results. 

\subsubsection{Variability due to stochasticity of training}

\begin{figure}
    \centering
    \includegraphics[width=10cm]{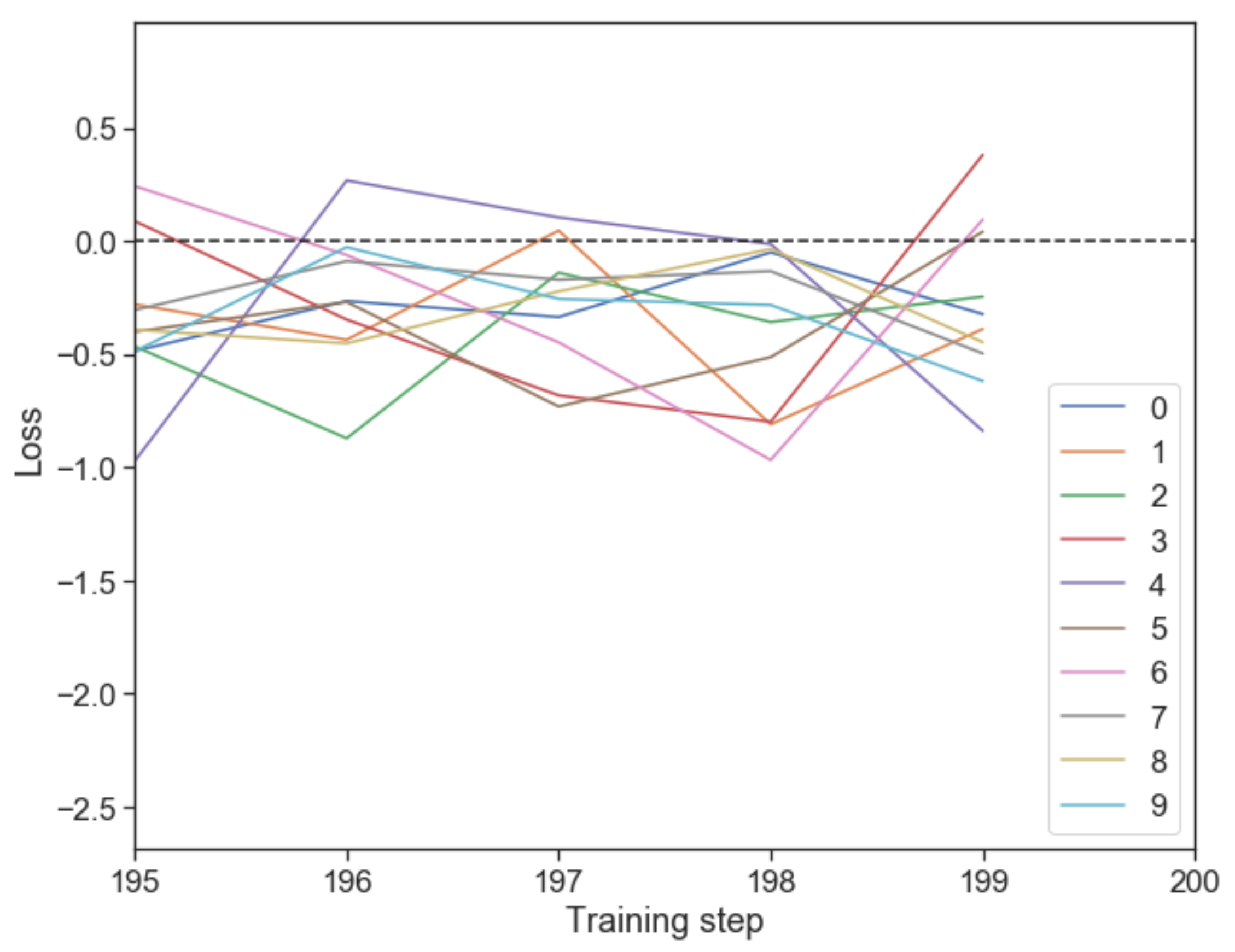}
    \caption{Network loss function for ten identical trainings of the same network with MCS 3000. The networks all show the same degree of loss fluctuation, suggesting that any trained network of this architecture would give similar results.}
    \label{fig:train}
\end{figure}

The final test considered whether random variations in the training of a given network could result in a particularly good or bad likelihood function. As such, variations would not be controllable with the user-defined parameters, so it is important to quantify this effect for reproducibility of the results. Figure~\ref{fig:train} shows the evolution of the network loss value when the same network (with MCS 3000) is trained ten times over the last 5 steps when the network is assumed to be fully trained. In all cases the loss function fluctuates between -1.0 and 0.5, with no one run showing a strong deviation. We therefore picked network 0 to use in our results.

\bibliographystyle{aasjournal}
\bibliography{ms.bib}{}

\end{document}